\documentclass[12pt]{article}
\usepackage{amsmath}
\usepackage{amssymb}

\renewcommand{\overline}[1]{\bar{#1}}
\makeatletter\@addtoreset{equation}{section}
\makeatother
\begin{document}
\begin{titlepage}
\begin{flushright}
TIT/HEP-591\\
HIP-2008-35/TH\\
November 2008\\
\end{flushright}
\vspace{0.5cm}
\begin{center}
{\Large \bf 

Instanton Calculus in R-R 3-form Background and Deformed $\mathcal{N} = 
2$ Super Yang-Mills Theory}
\lineskip .75em
\vskip1.0cm
{\large Katsushi Ito${}^{1}$, Hiroaki Nakajima${}^{2}$
and Shin Sasaki${}^{3}$ }
\vskip 2.5em
${}^{1}$ {\normalsize\it Department of Physics\\
Tokyo Institute of Technology\\
Tokyo, 152-8551, Japan}  \vskip 1.5em
${}^{2}$ {\normalsize\it BK21 Physics Research Division 
and Institute of Basic Science\\
Sungkyunkwan University\\
Suwon, 440-746, Korea
} \vskip 1.5em
${}^{3}$ {\normalsize\it Department of Physics
\\
Helsinki Institute of Physics, University of Helsinki\\
P.O.Box 64, FIN-00014, Finland 
} 
\vskip 4.5em
\end{center}
\begin{abstract}
We study the ADHM construction of instantons in 
$\mathcal{N} = 2$ supersymmetric  Yang-Mills theory 
deformed in constant Ramond-Ramond (R-R) 3-form field strength
background in type IIB superstrings.
We compare the deformed instanton effective action with 
the effective action
 of fractional D3/D$(-1)$ branes at the orbifold singularity of ${\bf
 C}^2/{\bf Z}_2$ in the  same R-R background.
We
find discrepancy between them 
at the second order in deformation parameters, which
comes from the coupling of the translational zero modes of 
the D$(-1)$-branes to the R-R background.
We improve the deformed action by adding a term with space-time
dependent gauge coupling.
Although the space-time action differs from the action in the $\Omega$-background, 
both actions lead to the same instanton equations of motion
at the lowest order in gauge coupling.
\end{abstract}
\end{titlepage}

\baselineskip=0.7cm

\tableofcontents

\section{Introduction}
Closed string background in superstring theories induces non-trivial
effects on D-branes, which are useful to study 
non-perturbative properties in 
supersymmetric gauge theories.
For example, constant NS-NS B-fields along D-branes
induce noncommutativity on the world-volume \cite{SeWi, ChuHo}.
Noncommutative instanton \cite{NeSc,Na} is a basic object for
studying the ADHM moduli space of instantons \cite{AtHiDrMa}, which resolves small 
instanton singularity.

Closed Ramond-Ramond (R-R) backgrounds also bring novel effects on the
D-branes.
In fact, constant self-dual graviphoton backgrounds are utilized to investigate
F-terms in 
supersymmetric gauge theories via closed/open string duality \cite{BeCeOoVa, AnGaNaTa, DiVa, OoVa}.
In this set-up, it is important to fix the scaling condition for the 
(self-dual) graviphoton
field strength ${\cal F}_{\alpha\beta}$, where $\alpha,\beta$ are spinor 
indices in four-dimensional space-time.
For example, in \cite{OoVa} 
the zero slope limit $\alpha'\rightarrow 0$ with fixed
$(2 \pi \alpha')^{- \frac{1}{2}} {\cal F}_{\alpha\beta}$ was considered.
On the other hand, the self-dual graviphoton background 
$\mathcal{F}_{\alpha\beta}$
with fixed
$(2 \pi \alpha')^{\frac{3}{2}} \mathcal{F}_{\alpha\beta}$ provides 
a non(anti)commutative deformation 
of $\mathcal{N} = 1$ superspace \cite{BeSe,DeGrNi,Se}.
${\cal N}=1$ supersymmetric gauge theories in non(anti)commutative
superspace has been studied extensively (see \cite{Se,ArItOh} for example).
The instanton solution and its moduli space 
are also deformed by non(anti)commutativity \cite{ArTaWa,BiFrPeLe}.
In \cite{BiFrPeLe}, the instanton is realized in D3/D$(-1)$-brane system at the
singularity on the orbifold ${\bf R}^6/{\bf Z}_2\times {\bf Z}_2$ in the
graviphoton background.

Non(anti)commutative ${\cal N}=1$ superspace is generalized to ${\cal
N}=2$ extended superspace, which admits the singlet and non-singlet type
of deformations \cite{IvLeZu}. 
Supersymmetric gauge theory on non(anti)commutative ${\cal N}=2$
harmonic superspace \cite{IvLeZu, FeIvSoZu,ArItOh2,CaIvLeQu,BuIvLeSaZu}
can be realized on D3-branes at the singularity in 
the orbifold ${\bf C}^2/{\bf Z}_2$ in the R-R 5-form background
${\cal F}^{\alpha\beta IJ}$ with the same scaling condition as in 
${\cal N}=1$ non(anti)commutative case.
Here  $I, J=1,2$ are
$SU(2)_R$ R-symmetry  indices.
It has been shown in \cite{ItSa} that 
symmetric-symmetric (S,S) type field strength ${\cal
F}^{(\alpha\beta)(IJ)}$ corresponds to the non-singlet deformation.
An antisymmetric-antisymmetric (A,A) type field strength
 ${\cal F}^{[\alpha\beta][IJ]}$
is expected to correspond to the singlet deformation
\cite{FeIvSoZu}.
The deformed instanton equation with some special deformation parameters
was discussed in \cite{ItNa}. 
Prepotential of non(anti)commutative gauge theory with singlet
deformation was also discussed in \cite{BuIvLeSaZu}.
Using string theory technique,
further extension to the $\mathcal{N} = 4$ gauge theory in the R-R
5-form graviphoton background was investigated  \cite{ItKoSa},
but their instanton solutions are not yet studied so far.
Recently its gravity dual has been proposed in \cite{ChDaSm}.

In superstring theory there are R-R
backgrounds with various rank. 
In type IIB theory, for example, there are R-R 3-forms (and its dual),
which correspond to the backgrounds ${\cal F}^{(\alpha\beta)[AB]}$ and
${\cal F}^{[\alpha\beta](AB)}$, denoted as (S,A) and (A,S) type deformations
\cite{ItKoSa}.
Here $A,B=1,\cdots,4$ are $SU(4)_R$ R-symmetry indices.
By orbifolding ${\bf C}^2/{\bf Z}_2$, we can introduce deformation of 
${\cal N}=2$ theory.
For the (S,A)-type deformation, the field strengths become
${\cal F}^{(\alpha\beta)(IJ)}$ and 
${\cal F}^{(\alpha\beta)[I'J']}$ ($I',J'=3,4$).
These deformations
cannot be realized in terms of non(anti)commutative 
superspace, but have interesting non-perturbative effects.

Recently, in \cite{BiFrFuLe} the low-energy effective action of
a system of fractional D3 and D$(-1)$-branes  was studied in the 
(S,A)-type background
 with fixed $(2 \pi \alpha')^{\frac{1}{2}}
 \mathcal{F}^{(\alpha\beta)[IJ]}$
and $(2 \pi \alpha')^{\frac{1}{2}} \mathcal{F}^{(\alpha\beta)[I'J']}$. 
They observed that the effective action of a fractional D3/D$(-1)$ system
agrees with the instanton effective actions of gauge theory in the
 $\Omega$-background \cite{LoMaNe} 
by identifying the R-R 3-form field strengths with 
the $\Omega$-background.
The instanton effective action in the $\Omega$ background plays an
important role to obtain the closed form of the prepotential in ${\cal
N}=2$ supersymmetric gauge theory with help of the localization
technique \cite{LoMaNe, Ne, NeOk, Shadchin, NaYo}.
Since the fractional D3/D$(-1)$ system in the R-R 3-form background
provides a simple string setup, it is important to study the relation between the R-R 3-form background 
and the $\Omega$-background in viewpoint of application to more general 
system.
In a previous paper \cite{ItNaSa},
 we studied deformation of ${\cal N}=2$ and $4$ super
Yang-Mills theories in the (S,A) or (A,S) type R-R 3-form background
\footnote{
In \cite{BiFrFuLe} the ${\cal N}=2$ deformed Lagrangian with ${\cal
F}^{(\alpha\beta)[I'J']}=0$ was obtained.
}.
It would be natural to expect that the 
deformed ${\cal N}=2$ gauge theory gives the effective action of 
D$(-1)$-branes in the R-R background.
However, there are some subtleties to identify both theories.
The deformed action of the D3-branes in the R-R 3-form background is
rather different from that of gauge theory in the $\Omega$-background.
Gauge theories deformed in the constant R-R 3-form
background 
have manifest translational symmetry, but the
$\Omega$-background metric contains space-time coordinates explicitly and 
translational invariance is lost.

The aim of this paper is to study the relation between 
${\cal N}=2$ super Yang-Mills theory deformed in the R-R 3-form
background
and the fractional D3/D$(-1)$ effective 
action.
We will solve the instanton equations in deformed theory using the ADHM
construction \cite{DoHoKhMa} 
up to the second order 
in the deformation parameter.
We then construct the instanton effective action from the field theory
and compare it with that obtained from the string theory.
We will see that discrepancy arises at the second order in the
deformation parameter, which comes from the absence of coupling
of translational zero modes to the R-R background
in the gauge theory side.
When we want to reproduce this coupling as an instanton solution, we
need to add one term to the deformed action at the second order.
The improved action has the space-time dependent gauge coupling,
which is similar to that in the $\Omega$-background. 
But two actions are shown to be different. However, they 
have the same instanton equations of motion
at the lowest order in gauge coupling and give the same instanton
effective action.

This analysis can be generalized into the ${\cal N}=4$ super Yang-Mills
theory in the R-R 3-form background and ${\cal N}=4$ version of the
$\Omega$-background.
This subject will be discussed in the next paper \cite{ItNaSaSa}.

The organization of this paper is as follows.
In section 2, we introduce four-dimensional (S,A)-deformed $\mathcal{N} = 
2$ $U(N)$ super Yang-Mills action defined on (fractional) D3-branes at 
the singularity of the orbifold $\mathbf{C}^2/\mathbf{Z}_2$.
The instanton equation is obtained and solved via the ADHM construction.
We calculate 
the instanton effective action for the self-dual solution and compare 
this result with the D3/D$(-1)$-branes result \cite{BiFrFuLe}.
However, once we introduce a term which breaks translational symmetry of the deformed action, 
both results agree even at the second order.

In section 3, the relation between the (S,A)-deformed super 
Yang-Mills theory and the $\Omega$-background is discussed.
Section 4 is devoted to conclusions and discussions.
We make a comment on 
the mass term deformation of the instanton effective action induced 
by the (A,S)-type background.
A brief introduction to the ADHM construction of instantons is 
presented in appendix A.
A detailed calculation of the instanton effective action can be found in 
appendix B.

\section{Instanton calculus in the (S,A)-deformed $\mathcal{N}=2$ super Yang-Mills theory}
In this section, we discuss four-dimensional $\mathcal{N} = 2$ 
$U(N)$ super Yang-Mills theory deformed by the (S,A)-type R-R 3-form 
background \cite{ItNaSa} 
and calculate the instanton solution and the instanton effective action. 
$\mathcal{N} = 2$ $U(N)$ super Yang-Mills theory is described by
gauge fields $A_\mu$ ($\mu=1,2,3,4$), 
complex scalars $\varphi$, $\bar{\varphi}$ and 
Weyl fermions $\Lambda_{\alpha}^{I}$ and
$\bar{\Lambda}^{\dot{\alpha}}_{I}$ ($I=1,2$), 
which belong to the adjoint representation of gauge group $U(N)$.
We denote $T^m$ as the basis of $U(N)$ generators normalized as
${\rm Tr}(T^m T^n)=\kappa\delta^{mn}$ with constant 
$\kappa$. The Lagrangian is given by 
\begin{align}
\mathcal{L}_{0} &= 
\frac{1}{\kappa} \mathrm{Tr} \left[ -\frac{1}{4} F_{\mu \nu}
 F^{\mu \nu} + \frac{i \theta g^2}{32 \pi^2} F_{\mu \nu}
 \tilde{F}^{\mu \nu} - D_{\mu} \varphi D^{\mu} \bar{\varphi} - \frac{1}{2} 
g^{2} [\varphi, \bar{\varphi}]^2   \right. \nonumber \\
& \qquad \qquad \qquad \left. - i \Lambda^{I \alpha} 
			  (\sigma^{\mu})_{\alpha \dot{\beta}} D_{\mu} 
			  \bar{\Lambda}_{I}^{\dot{\beta}}
+ \frac{i}{\sqrt{2}}g\, \Lambda^{I} [\bar{\varphi}, \Lambda_{I}] -
\frac{i}{\sqrt{2}}g\, \bar{\Lambda}_{I} [\varphi, \bar{\Lambda}^{I}] 
\right]. \label{N2SYM} 
\end{align}
Here $F_{\mu \nu} = \partial_{\mu} A_{\nu} - \partial_{\nu} A_{\mu}
+ i g [A_{\mu}, A_{\nu}]$ is the gauge field strength, $g$ is  
the gauge coupling constant 
and $D_{\mu} * = \partial_{\mu} * + i g [A_{\mu}, *]$ 
is a gauge covariant derivative. 
We also define $\sigma_\mu=(i\tau^1,i\tau^2,i\tau^3,1)$ and
$\bar{\sigma}_\mu=(-i\tau^1,-i\tau^2,-i\tau^3,1)$,
where $\tau^i$ ($i=1,2,3$) are the Pauli matrices. $\theta$ is a 
theta angle 
and $\tilde{F}_{\mu \nu} = \frac{1}{2} \varepsilon_{\mu \nu \rho 
\sigma} F^{\mu \nu}$. 
This theory is the low-energy effective theory of $N$ (fractional) D3-branes 
on $\mathbf{C}\times\mathbf{C}^2/\mathbf{Z}_2$, where 
the D3-branes are located in the fixed point of the orbifold \cite{BeDiFrLeMa-BeDiMa}.

We now introduce the (S,A)-type R-R 3-form 
$\mathcal{F}^{(\alpha \beta) [AB]}$. 
After ${\bf Z}_2$ orbifolding, 
the surviving components
are $\mathcal{F}^{(\alpha \beta)12}$ and $\mathcal{F}^{(\alpha \beta)34}$, 
from which 
we define $\mathcal{N} = 2$ deformation parameters as 
$C^{\alpha \beta} = 4 \sqrt{2} \pi (2 \pi \alpha')^{\frac{1}{2}}
\mathcal{F}^{(\alpha \beta) 12}$, $ 
\bar{C}^{\alpha \beta} = 4 \sqrt{2} \pi(2 \pi \alpha')^{\frac{1}{2}} 
\mathcal{F}^{(\alpha \beta) 34}$. 

We also use the notation 
$C^{\mu\nu} \equiv 
\varepsilon_{\beta \gamma} (\sigma^{\mu \nu})_{\alpha} {}^{\gamma} 
C^{\alpha \beta}$ and $\bar{C}^{\mu\nu} \equiv 
\varepsilon_{\beta \gamma} (\sigma^{\mu \nu})_{\alpha} {}^{\gamma} 
\bar{C}^{\alpha \beta}$ where $\sigma^{\mu \nu} = \frac{1}{4} (
\sigma^{\mu} \bar{\sigma}^{\nu} - \sigma^{\nu} \bar{\sigma}^{\mu} )$.
$C^{\mu\nu}$ corresponds to the self-dual graviphoton field strength in 
$\mathcal{N} = 2$ supergravity multiplet while $\bar{C}^{\mu\nu}$ 
corresponds to the self-dual background of the vector multiplet 
\cite{BiFrFuLe}. 
The deformed Lagrangian up to the second order in the deformation 
parameter is \cite{ItNaSa}
\begin{equation}
\mathcal{L}=\mathcal{L}_{0}+{}
\mathcal{L}_{C},
\label{N2Lagrangian}
\end{equation}
where the second 
term $\mathcal{L}_{C}$ in (\ref{N2Lagrangian}) 
is the interaction term 
obtained from the computation of disk amplitudes of open strings 
in the R-R 3-form background; 
\begin{align}
\mathcal{L}_{C}&= 
 \frac{1}{\kappa} \mathrm{Tr} \left[
ig(C^{\mu \nu }\bar{\varphi}+\bar{C}^{\mu \nu}\varphi)F_{\mu \nu}
+ \frac{i}{\sqrt{2}} g \Lambda_{\alpha} {}^{I} \Lambda_{\beta I} 
\bar{C}^{(\alpha \beta)}
{}+\frac{1}{2}g^{2}(C^{\mu \nu }\bar{\varphi}+\bar{C}^{\mu \nu}\varphi)^2
\right].
\end{align}

We study the instanton solution of the deformed theory 
based on the Euclidean action.
The bosonic part relevant to the gauge instanton
is written in the perfect square form $S'$ as 
\begin{align}
S'
&=\!\int\!d^{4}x\,\frac{1}{\kappa}\mathrm{Tr}\biggl[
\frac{1}{2}\bigl(F_{\mu\nu}^{(+)}
-ig(C^{\mu\nu}\bar{\varphi}+\bar{C}^{\mu\nu}\varphi)\bigr)^{2}\Biggr]
+\biggl(- \frac{8\pi^{2}}{g^{2}}+i\theta\biggr)k
\notag\\
&=\!\int\!d^{4}x\,\frac{1}{\kappa}\mathrm{Tr}\biggl[
\frac{1}{2}\bigl(F_{\mu\nu}^{(-)}\bigr)^{2}
- ig(C^{\mu\nu}\bar{\varphi}+\bar{C}^{\mu\nu}\varphi)F^{(+)}_{\mu\nu}
- \frac{g^{2}}{2}(C^{\mu\nu}\bar{\varphi}+\bar{C}^{\mu\nu}\varphi)^{2}
\Biggr]+\biggl(\frac{8\pi^{2}}{g^{2}}+i\theta\biggr)k.
\label{square}
\end{align}
where $F^{(\pm)}_{\mu \nu} = \frac{1}{2} (F_{\mu \nu} \pm \tilde{F}_{\mu 
\nu})$.
The instanton number $k$ is defined by
\begin{equation}
k = \frac{g^2}{32 \pi^2} \int \! d^4 x \ \frac{1}{\kappa}
\mathrm{Tr} F_{\mu \nu} \tilde{F}^{\mu \nu}.
\end{equation}
We then obtain the self-dual and 
the anti-self-dual equations.
\begin{align}
F^{(-)}_{\mu \nu} &= 0, \label{instanton_solution} \\
F^{(+)}_{\mu \nu} - i g (C^{\mu \nu} \bar{\varphi} + \bar{C}^{\mu \nu} 
\varphi) &= 0.
\label{anti-instanton_solution}
\end{align}
A solution corresponding to the equation (\ref{instanton_solution}) 
is called the self-dual solution while the one to the equation 
(\ref{anti-instanton_solution}) is the anti-self-dual solution.
The other fields satisfy the equation of 
motion in the (anti-)self-dual background. 
{}From the Lagrangian (\ref{N2Lagrangian}), the equations of motion
are derived as 
\begin{eqnarray}
& & D^2 \bar{\varphi} - i \sqrt{2}  g \overline{\Lambda}_I
\overline{\Lambda}^I - g^2 \Bigl[ \bar{\varphi}, [\varphi, \bar{\varphi}]
\Bigr] + i g F_{\mu \nu} \bar{C}^{\mu \nu} + 
g^2 \varphi \bar{C}_{\mu \nu} \bar{C}^{\mu \nu} + g^2
\bar{\varphi} C_{\mu \nu} \bar{C}^{\mu \nu}
= 0, \nonumber \\
& & D^2 \varphi + i \sqrt{2} g \Lambda^I \Lambda_I 
- g^2 \Bigl[ \varphi, [\bar{\varphi}, \varphi] \Bigr]
+ i g F_{\mu \nu} C^{\mu \nu} + g^2 
\bar{\varphi} C_{\mu \nu} C^{\mu \nu} + g^2
\varphi C_{\mu \nu} \bar{C}^{\mu \nu} = 0, \nonumber \\
& & (\sigma^{\mu})_{\alpha \dot{\beta}} D_{\mu} \overline{\Lambda}_I
{}^{\dot{\beta}} + \sqrt{2} g [\bar{\varphi}, \Lambda_{I \alpha}]
+ \sqrt{2} g\, \bar{C}_{\alpha \beta} \Lambda^{\beta}_I
= 0, \nonumber\\
& & (\bar{\sigma}^{\mu})^{\dot{\alpha} \beta} D_{\mu} \Lambda^I {}_{\beta}
- \sqrt{2} g [\varphi, \overline{\Lambda}^{I 
\dot{\alpha}}] = 0, \nonumber\\
& & D_{\mu} \left( F^{\mu \nu} 
- 2i g \bar{\varphi} C^{\mu \nu} 
- 2i g \varphi \bar{C}^{\mu \nu}
\right) \nonumber \\
& & \qquad \qquad - i g [\varphi, D^{\nu} \bar{\varphi}] 
- i g [\bar{\varphi}, D^{\nu} \varphi]
- g (\sigma^{\nu})_{\alpha \dot{\beta}}
\{\Lambda^{I \alpha}, \overline{\Lambda}_I {}^{\dot{\beta}} \} = 0.
\label{eom}
\end{eqnarray}
First we consider the case where the vacuum expectation values (VEVs) 
of the scalar fields are zero. In this case, 
we find some exact solutions. 
For example, in the case of $\bar{C}^{\mu \nu} = 0$, the Dirac equation for 
the fermion $\bar{\Lambda}_{\dot{\alpha}}$ has no zero mode in the self-dual background.
We can set $\bar{\Lambda} = 0 $. Then 
the equation of motion for $\bar{\varphi}$ becomes 
\begin{equation}
D^2 \bar{\varphi} - g^2 \Bigl[ \bar{\varphi}, [\varphi, \bar{\varphi}]
\Bigr] 
= 0,
\end{equation}
from which $\bar{\varphi} = 0$ is found to be an exact solution.
Therefore $\bar{\varphi} = \bar{\Lambda} = 0$ 
is shown to be an exact solution. 
Then the equation of motion for the other fields becomes 
\begin{align}
&F_{\mu\nu}^{(-)}=0,\label{self01} \\
&(\bar{\sigma}^{\mu})^{\dot{\alpha}\beta}D_{\mu}\Lambda^{I}_{\beta}=0, \\
&D^{2}\varphi+i\sqrt{2}g\Lambda^{I}\Lambda_{I}+iC^{\mu\nu}F_{\mu\nu}=0.
\label{self03}
\end{align}
The equations \eqref{self01}--\eqref{self03} are solved by 
the ADHM construction \cite{DoHoKhMa} (see appendix \ref{revADHM})
for any instanton number $k$ as
\begin{align}
A_{\mu} &= -i \overline{U} \partial_{\mu} U, 
\\
\Lambda^{I}_{\alpha} &= 
\overline{U} (\mathcal{M}^I f \bar{b}_{\alpha} - b_{\alpha} f 
\overline{\mathcal{M}}^I) U, 
\\
\varphi &= - i \frac{\sqrt{2}}{4} \epsilon_{IJ}
\bar{U} \mathcal{M}^I f \bar{\mathcal{M}}^J U + \bar{U}
\left(
\begin{array}{cc}
0 & 0 \\
0 & \chi \mathbf{1}_{2} + \mathbf{1}_{k} C
\end{array}
\right)
U. \label{novev}
\end{align}
Here $U$ is the $(N+2k)\times N$ matrix which satisfies 
$\bar{\Delta}^{\dot{\alpha}}U=0$ with 
the $(N+2k)\times 2k$ matrix 
\begin{equation}
\Delta_{\dot{\alpha}}=a_{\dot{\alpha}}
+b^{\beta}(\sigma_{\mu})_{\beta\dot{\alpha}}x^{\mu}
=\binom{w_{\dot{\alpha}}}{(a'+x)_{\alpha\dot{\alpha}}}, 
\quad x_{\alpha\dot{\alpha}}=(\sigma_{\mu})_{\alpha\dot{\alpha}}x^{\mu},
\end{equation}
where the parameters 
$a'_{\mu}=\frac{1}{2}(\bar{\sigma}_{\mu})^{\dot{\alpha}\alpha}
a'_{\alpha\dot{\alpha}}$ and $w_{\dot{\alpha}}$ satisfy the ADHM constraints 
\begin{equation}
(\vec{\tau})^{\dot{\alpha}}_{~\dot{\beta}}
(\bar{w}^{\dot{\beta}}w_{\dot{\alpha}}
+\bar{a}^{\prime\dot{\beta}\alpha}a'_{\alpha\dot{\alpha}})=0,\quad 
a'_{\mu}=\bar{a}'_{\mu}.
\label{ADHM}
\end{equation}
$\mathcal{M}^{I}=(\mu^{I}\ \mathcal{M}^{\prime I}_{\alpha})^{T}$ is the 
$(N+2k)\times k$ constant Grassmann-odd matrix 
which satisfies the fermionic ADHM constraints
\begin{equation}
\bar{\mu}^{I}w_{\dot{\alpha}}+\bar{w}_{\dot{\alpha}}\mu^{I}
+[\mathcal{M}^{\prime\alpha I},a'_{\alpha\dot{\alpha}}]=0,
\quad \mathcal{M}^{\prime I}_{\alpha}=\bar{\mathcal{M}}^{\prime I}_{\alpha}.
\label{fADHM}
\end{equation}
The parameters 
$a'_{\alpha\dot{\alpha}}$, $w^{}_{\dot{\alpha}}$, 
$\mathcal{M}^{\prime I}_{\alpha}$ and $\mu^{I}$ are called ADHM moduli. 
$C$ in \eqref{novev} is the $2\times 2$ matrix
of which components are
$C_{\alpha}{}^{\beta} = 
\frac{1}{2}(\sigma_{\mu\nu})_{\alpha}{}^{\beta}C^{\mu\nu}$. 
The $k\times k$ matrix 
$\chi$ obeys the following equation such that \eqref{novev} is a solution of 
\eqref{self03}:
\begin{align}
\boldsymbol{L} \chi &= 
i \frac{\sqrt{2}}{4} \epsilon_{IJ}
\overline{\mathcal{M}}^I \mathcal{M}^J 
+ C^{\mu \nu} [a'_{\mu}, a'_{\nu}], 
\end{align}
where the operator $\boldsymbol{L}$ is defined by
\begin{equation}
\boldsymbol{L} * = \frac{1}{2} 
\bigl\{ \bar{w}^{\dot{\alpha}} w_{\dot{\alpha}}, * \bigr\}
+ \Bigl[a'_{\mu}, [a'^{\mu}, *]\Bigr]. 
\label{L}
\end{equation}

In the case of $C^{\mu\nu}=0$, 
a solution of 
\begin{align}
&F_{\mu\nu}^{(+)}=0, \label{antiself01}\\
&(\sigma^{\mu})_{\alpha\dot{\beta}}D_{\mu}\bar{\Lambda}^{\dot{\beta}}_{I}=0, \\
&D^{2}\bar{\varphi}-i\sqrt{2}g\bar{\Lambda}_{I}\bar{\Lambda}^{I}
=0, \\
&\Lambda_{\alpha}^{I}=0,\quad \varphi=0 
\label{antiself04}
\end{align}
satisfies 
the equations of motion. 
In this case, the solution is independent of $\bar{C}^{\mu\nu}$ 
because the self-duality of $\bar{C}^{\mu\nu}$ leads to
$\bar{C}^{\mu\nu}F^{(-)}_{\mu\nu}=0$ in (\ref{eom}).
Therefore 
the anti-self-dual solution is not deformed by $\bar{C}^{\mu\nu}$ when $C^{\mu\nu}=0$.

Nextly we consider the case where both $C^{\mu\nu}$ and 
$\bar{C}^{\mu\nu}$ are nonzero and 
where the adjoint scalar fields $\varphi, \bar{\varphi}$ have nonzero VEVs.
In this case, 
we should consider the constrained instanton solution 
(see \cite{DoHoKhMa} for a review). 
We solve the equations of motion perturbatively in the gauge coupling $g$.
The expansion in $g$ gives reliable results 
when the VEVs $\phi=\langle\varphi\rangle$ and 
$\bar{\phi}=\langle\bar{\varphi}\rangle$ are large.
Then in the self-dual background 
the classical action $S$ is expanded as 
\begin{equation}
S = \frac{8 \pi^2 k}{g^2} + i k \theta + g^0 S^{(0)}_{\mathrm{eff}} +
\mathcal{O} (g^2). \label{classical_g-expansion}
\end{equation}
$S^{(0)}_{\mathrm{eff}}$ is called the instanton effective action. 
The instanton effective action in the anti-self-dual background is 
also defined similarly.
$S^{(0)}_{\mathrm{eff}}$ is expressed by the ADHM moduli parameters 
by plugging the constrained instanton 
solution into the action.

In the next subsections, we investigate the constrained instanton solutions. 
We will discuss the solution for the self-dual and the anti-self-dual 
cases separately.

\subsection{Anti-self-dual case}
For the anti-self-dual case 
(\ref{anti-instanton_solution}), the solution is expanded 
in the gauge coupling $g$ as 
\begin{eqnarray}
A_{\mu} &=& g^{-1} A^{(0)}_{\mu} + 
g A_{\mu}^{(1)} + \cdots, \label{N2ASD1} \\
\Lambda^I &=& g^{ \frac{1}{2}} \Lambda^{(0)I} + 
g^{\frac{5}{2}} \Lambda^{(1)I} + \cdots, \label{N2ASD2} \\
\bar{\Lambda}_I &=& g^{- \frac{1}{2}} \bar{\Lambda}^{(0)}_I
+ g^{\frac{3}{2}} \bar{\Lambda}^{(1)}_I + \cdots, \label{N2ASD3} \\
\varphi &=& g^0 \varphi^{(0)} + g^2 \varphi^{(1)} + \cdots, 
\label{N2ASD4} \\
\bar{\varphi} &=& g^0 \bar{\varphi}^{(0)} + g^2 
\bar{\varphi}^{(1)} + \cdots. \label{N2ASD5}
\end{eqnarray}
The equations of motion (\ref{eom}) for the fields at the leading order become 
\begin{eqnarray}
& & F^{(0)(+)}_{\mu \nu} = 0, \\
& & \nabla^2 \bar{\varphi}^{(0)} - i \sqrt{2} 
\overline{\Lambda}^{(0)}_I \overline{\Lambda}^{(0)I} = 0, \\
& & \nabla^2 \varphi^{(0)} = 0,\\
& & (\sigma^{\mu})_{\alpha \dot{\beta}} \nabla_{\mu} 
\overline{\Lambda}^{(0) \dot{\beta}}_I = 0, \\
& & (\bar{\sigma}^{\mu})^{\dot{\alpha} \beta} \nabla_{\mu} 
\Lambda^{(0)I}_{\beta} - \sqrt{2} [\varphi^{(0)}, 
\overline{\Lambda}^{(0) I \dot{\alpha}}] = 0, \\
& & \nabla_{\mu} 
F^{(0) \mu \nu} 
= 0,
\end{eqnarray}
where $\nabla_{\mu}$ denotes the covariant derivative in the 
instanton background $\nabla_{\mu}=\partial_{\mu}+i[A_{\mu}^{(0)},\ast]$. 
These equations are not deformed.
The instanton effective 
action $S^{(0)}_{\mathrm{eff}}$ 
in (\ref{classical_g-expansion}) is 
evaluated as 
\begin{eqnarray}
S^{(0)}_{\mathrm{eff}} = 
\frac{1}{\kappa} \int \! d^4 x \ \mathrm{Tr} \left[
\nabla_{\mu} \varphi^{(0)} \nabla^{\mu} \bar{\varphi}^{(0)}
 -\frac{i}{\sqrt{2}} \overline{\Lambda}_I^{(0)} [\varphi^{(0)}, 
 \overline{\Lambda}^{(0)I} ]
\right], 
\label{antiinsteff}
\end{eqnarray}
which is not also deformed.

\subsection{Self-dual case}
For the self-dual case 
(\ref{instanton_solution}), we have the expansion
\begin{align}
A_{\mu} &= g^{-1} A^{(0)}_{\mu} + 
g A_{\mu}^{(1)} + \cdots, \label{N2SD1} \\
\Lambda^I &= g^{- \frac{1}{2}} \Lambda^{(0)I} + 
g^{\frac{3}{2}} \Lambda^{(1)I} + \cdots, \label{N2SD2} \\
\bar{\Lambda}_I &= g^{\frac{1}{2}} \bar{\Lambda}^{(0)}_I
+ g^{\frac{5}{2}} \bar{\Lambda}^{(1)}_I + \cdots, \label{N2SD3} \\
\varphi &= g^0 \varphi^{(0)} + g^2 \varphi^{(1)} + \cdots, 
\label{N2SD4} \\
\bar{\varphi} &= g^0 \bar{\varphi}^{(0)} + g^2 
\bar{\varphi}^{(1)} + \cdots. \label{N2SD5}
\end{align}
The equations of motion at the leading order are 
\begin{align}
& F^{(0)(-)}_{\mu \nu} = 0, \label{N2_leading_self-dual} \\
& \nabla^2 \bar{\varphi}^{(0)} + i F^{(0)}_{\mu \nu} \bar{C}^{\mu \nu}= 0, 
\label{eom2}\\
& \nabla^2 \varphi^{(0)} + i \sqrt{2}  \Lambda^{(0)I} 
\Lambda^{(0)}_I + i F^{(0)}_{\mu \nu} C^{\mu \nu} = 0, \\
& (\bar{\sigma}^{\mu})^{\dot{\alpha} \beta} \nabla_{\mu} 
\Lambda^{(0)I}_{\beta} = 0, \label{leading4}\\
& (\sigma^{\mu})_{\alpha \dot{\beta}} \nabla_{\mu}
\overline{\Lambda}^{(0)}_I {}^{\dot{\beta}} + \sqrt{2} 
[\bar{\varphi}^{(0)}, \Lambda^{(0)}_{I \alpha}] 
+ \sqrt{2} \Lambda^{(0)\beta} {}_{I} \bar{C}_{(\beta \alpha)}= 0, \\
& \nabla_{\mu} F^{(0) \mu \nu} 
= 0,
\label{leading_eq}
\end{align}
The equations \eqref{leading_eq} is automatically satisfied due to the self-dual condition 
(\ref{N2_leading_self-dual}).
Other equations \eqref{N2_leading_self-dual}--\eqref{leading4}
have been solved via the ADHM construction 
in the case of $\bar{C}^{\mu \nu} = 0$ \cite{BiFrFuLe}. 
For nonzero $C^{\mu\nu}$ and $\bar{C}^{\mu\nu}$ these are solved as
\begin{eqnarray}
A^{(0)}_{\mu} &=& -i \overline{U} \partial_{\mu} U, \label{sol1}\\
\Lambda^{(0)I}_{\alpha} &=& 
\overline{U} (\mathcal{M}^I f \bar{b}_{\alpha} - b_{\alpha} f 
\overline{\mathcal{M}}^I) U, \label{sol_gauge} \\
\varphi^{(0)} &=& - i \frac{\sqrt{2}}{4} \epsilon_{IJ}
\bar{U} \mathcal{M}^I f \bar{\mathcal{M}}^J U + \bar{U}
\left(
\begin{array}{cc}
\phi & 0 \\
0 & \chi \mathbf{1}_{2} + \mathbf{1}_{k} C
\end{array}
\right)
U, \label{sol_lambda}\\
\bar{\varphi}^{(0)} &=& \bar{U}
\left(
\begin{array}{cc}
\bar{\phi} & 0 \\
0 & \bar{\chi} \mathbf{1}_{2} + \mathbf{1}_{k} \bar{C}
\end{array}
\right)
U.\label{sol_adjoint2}
\end{eqnarray}
Here $\bar{C}$ is the $2\times 2$ matrix
of which components are
$\bar{C}_{\alpha}{}^{\beta} = 
\frac{1}{2}(\sigma_{\mu\nu})_{\alpha}{}^{\beta}\bar{C}^{\mu\nu}$. 
The $k\times k$ matrices $\chi$ and $\bar{\chi}$  
satisfy the equations
\begin{align}
\boldsymbol{L} \chi &= 
i \frac{\sqrt{2}}{4} \epsilon_{IJ}
\overline{\mathcal{M}}^I \mathcal{M}^J 
+\bar{w}^{\dot{\alpha}} \phi w_{\dot{\alpha}}
+ C^{\mu \nu} [a'_{\mu}, 
a'_{\nu}], \label{chi_const1} \\
\boldsymbol{L} \bar{\chi} &= 
\bar{w}^{\dot{\alpha}} \bar{\phi} w_{\dot{\alpha}}
+ \bar{C}^{\mu \nu} [a'_{\mu}, a'_{\nu}] \label{chi_const2}.
\end{align}

We note that we do not need to solve 
the equation of motion for $\bar{\Lambda}_{I}^{(0)\dot{\alpha}}$ explicitly. 
This is because 
contribution of $\bar{\Lambda}_{I}^{(0)\dot{\alpha}}$
to the 
action is just the subleading order 
in gauge coupling constant $g$. 
We also note 
that the solutions of the gauge field $A_{\mu}$ and 
Weyl fermion $\Lambda^{(0)I}_{\alpha}$ are not deformed by 
$C^{\mu\nu}$ and $\bar{C}^{\mu\nu}$ and the ADHM constraints 
\eqref{ADHM} and \eqref{fADHM} (see appendix \ref{revADHM}) 
do not change. 
This is contrasted with the case of $\mathcal{N} = 1$ non(anti)commutative 
deformed super Yang-Mills \cite{BiFrPeLe, ArTaWa} in which the bosonic ADHM 
constraints \eqref{ADHM} 
are changed due to the non-zero graviphoton 
background while fermionic constraints \eqref{fADHM} remain unchanged.

Now let us evaluate the instanton effective action in the 
self-dual instanton background and write down it
in terms of the ADHM moduli.
Some formulae are proved in appendix \ref{calc}.
By substituting the expansion (\ref{N2SD1})-(\ref{N2SD5}) into 
the classical action, the instanton effective action 
is given by
\begin{align}
S^{(0)}_{\mathrm{eff}} &= 
\frac{1}{\kappa} \int \! d^4 x \ \mathrm{Tr} \left[ 
\nabla_{\mu} \varphi^{(0)} \nabla^{\mu} \bar{\varphi}^{(0)} 
- \frac{i}{\sqrt{2}}
\Lambda^{(0)I} [\bar{\varphi}, \Lambda^{(0)}_I] - i 
\bar{\varphi}^{(0)} F^{(0)}_{\mu \nu} C^{\mu \nu} \right. 
\nonumber \\
& \qquad \qquad \qquad \qquad \qquad \left. - i \varphi^{(0)} 
F^{(0)}_{\mu \nu} \bar{C}^{\mu \nu} 
- \frac{i}{\sqrt{2}}   \Lambda_{\alpha}^{(0)I} \Lambda_{\beta I}^{(0)} 
\bar{C}^{(\alpha \beta)}
\right]. 
\label{instanton_effective_action}
\end{align}
From the equation of motion \eqref{eom2}, 
the first and the fourth terms in \eqref{instanton_effective_action} become 
the total derivative, 
\begin{align}
\frac{1}{\kappa} \int \! d^4 x \ \mathrm{Tr} \left[ 
\nabla_{\mu} \varphi^{(0)} \nabla^{\mu} \bar{\varphi}^{(0)} 
- i \varphi^{(0)} F^{(0)}_{\mu \nu} \bar{C}^{\mu \nu}
\right]
=
\int \! d^4 x \ \frac{1}{\kappa} \mathrm{Tr} \Bigl[ 
 \partial_{\mu} \bigl( \varphi^{(0)} \nabla^{\mu} \bar{\varphi}^{(0)} \bigr)
\Bigr],
\end{align}
which is evaluated by the value of $\nabla_{\mu}\varphi^{(0)}$ at infinity as
\begin{align}
\int \! d^4 x \ \frac{1}{\kappa} \mathrm{Tr} \Bigl[ 
 \partial_{\mu} \bigl( \varphi^{(0)} \nabla^{\mu} \bar{\varphi}^{(0)} \bigr)
\Bigr]
&= \frac{1}{\kappa}
\lim_{|x|\to\infty}2\pi^{2}|x|^{2}x^{\mu}\mathrm{Tr}\bigl[
\varphi^{(0)}\nabla_{\mu}\bar{\varphi}^{(0)}\bigr]\notag\\
&=\frac{4\pi^{2}}{\kappa}\,\mathrm{tr}_{k}\biggl[\frac{1}{2}\bar{w}^{\dot{\alpha}}
(\bar{\phi}\phi+\phi\bar{\phi})w_{\dot{\alpha}}
-\bar{w}^{\dot{\alpha}}\phi w_{\dot{\alpha}}\bar{\chi}
\biggr]. \label{first}
\end{align}
Here $\mathrm{tr}_{k}$ denotes the trace for instanton indices. 
The second and the fifth terms 
in \eqref{instanton_effective_action}
are calculated as 
(see appendix \ref{calc})
\begin{align}
&\int\!d^{4}x\,\frac{1}{\kappa}\mathrm{Tr}\biggl[
-\frac{i}{\sqrt{2}}
\Lambda^{(0)\alpha I}\bigl[\bar{\varphi}^{(0)},\Lambda_{\alpha I}^{(0)}\bigr]
- \frac{i}{\sqrt{2}}\bar{C}^{(\alpha\beta)}\Lambda_{\alpha}^{(0)I}
\Lambda_{\beta I}^{(0)}\biggr]\notag\\
&\qquad\qquad=
\frac{1}{\kappa} \sqrt{2}\pi^{2}i \epsilon_{IJ}
\mathrm{tr}_{k}\biggl[\bar{\mu}^{I}\bar{\phi}\mu^{J}
-\bar{\mathcal{M}}^{I}\mathcal{M}^{J}\bar{\chi}
+\frac{1}{2}\bar{C}^{(\alpha\beta)}\mathcal{M}^{\prime I}_{\alpha}
\mathcal{M}^{\prime J}_{\beta}\biggr].
\label{secondthird}
\end{align}
The third term is 
\begin{align}
&\int\!d^{4}x\,\frac{1}{\kappa}\mathrm{Tr}\Bigl[
- i \bar{\varphi}^{(0)} F^{(0)}_{\mu \nu} {C}^{\mu \nu}
\Bigr]=
\frac{\pi^{2}}{\kappa}\mathrm{tr}_{k}\Bigl[
-4C^{\mu\nu}[a'_{\mu},a'_{\nu}]\bar{\chi}
+C^{\mu\nu}\bar{C}_{\mu\nu}
\bar{w}^{\dot{\alpha}}w_{\dot{\alpha}}
\Bigr].
\label{fourth}
\end{align}
Finally $S^{(0)}_{\mathrm{eff}}$ becomes
\begin{align}
S^{(0)}_{\mathrm{eff}}&=
\frac{4\pi^{2}}{\kappa}\mathrm{tr}_{k}\biggl[
-\Bigl(\bar{w}^{\dot{\alpha}} \bar{\phi} w_{\dot{\alpha}}
+ \bar{C}^{\mu \nu} [a'_{\mu}, a'_{\nu}]\Bigr)
\boldsymbol{L}^{-1}\Bigl(i\frac{\sqrt{2}}{4}\epsilon_{IJ}
\bar{\mathcal{M}}^{I}\mathcal{M}^{J}
+\bar{w}^{\dot{\alpha}}\phi w_{\dot{\alpha}}
+C^{\mu\nu}[a'_{\mu},a'_{\nu}]\Bigr)
\notag\\
&\qquad\qquad\quad{}
+i\frac{\sqrt{2}}{4}\epsilon_{IJ}
\bar{\mu}^{I}\bar{\phi}\mu^{J}+\frac{1}{2}\bar{w}^{\dot{\alpha}}
(\bar{\phi}\phi+\phi\bar{\phi})w_{\dot{\alpha}}
-i\frac{\sqrt{2}}{8}\bar{C}^{(\alpha\beta)}\epsilon_{IJ}
\mathcal{M}^{\prime I}_{\alpha}\mathcal{M}^{\prime J}_{\beta}
\notag\\
&\qquad\qquad\quad{}
+\frac{1}{4}C^{\mu\nu}\bar{C}_{\mu\nu}\bar{w}^{\dot{\alpha}}w_{\dot{\alpha}}
\biggr].
\label{insteff}
\end{align}
This action can be also 
obtained from the following action by integrating over the auxiliary fields 
$\chi$, $\bar{\chi}$, $\bar{\psi}^{\dot{\alpha}}_{I}$ and $\vec{D}$
\begin{align}
S^{(0)}_\mathrm{eff}&= \frac{2\pi^{2}}{\kappa}\mathrm{tr}_{k}\biggl[
-2\Bigl([\bar{\chi},a'_{\mu}] - \bar{C}_{\mu\nu}a^{\prime \nu}\Bigr)
\Bigl([\chi,a^{\prime \mu}] - C^{\mu\rho}a'_{\rho}\Bigr)\notag\\
&\quad{}+(\bar{\chi}\bar{w}^{\dot{\alpha}}-\bar{w}^{\dot{\alpha}}\bar{\phi})
(w_{\dot{\alpha}}\chi-\phi w_{\dot{\alpha}})
+(\chi\bar{w}^{\dot{\alpha}}-\bar{w}^{\dot{\alpha}}\phi)
(w_{\dot{\alpha}}\bar{\chi}-\bar{\phi}w_{\dot{\alpha}})\notag\\
&\quad{}-i\frac{\sqrt{2}}{2}\bar{\mu}^{I}\epsilon_{IJ}(\mu^{J}\bar{\chi}
-\bar{\phi}\mu^{J})
-i\frac{\sqrt{2}}{4}\mathcal{M}^{\prime\alpha I}\epsilon_{IJ}
\Bigl([\bar{\chi},\mathcal{M}_{\alpha}^{\prime J}]
-\bar{C}_{(\alpha\beta)}\mathcal{M}^{\prime\beta J}\Bigr)
\notag\\
&\quad{}
+\frac{1}{2}C^{\mu\nu}\bar{C}_{\mu\rho}(\delta_{\nu}{}^{\rho}
\bar{w}^{\dot{\alpha}}w_{\dot{\alpha}}
+4a'_{\nu}a^{\prime \rho})\biggr]+S_{\mathrm{ADHM}},
\label{insteff2}
\end{align}
where $S_{\mathrm{ADHM}}$ 
contains the Lagrange multipliers $\bar{\psi}^{\dot{\alpha}}_I, \vec{D}$ 
associated with the ADHM constraints \eqref{ADHM} and \eqref{fADHM} 
by its equation of motion. 
It is given by 
\begin{align}
S_{\mathrm{ADHM}}&=\frac{4\pi^2}{\kappa}\mathrm{tr}_k
\biggl[
-i\bar{\psi}^{\dot{\alpha}}_{I}
\Bigl(\bar{\mu}^{I}w_{\dot{\alpha}}+\bar{w}_{\dot{\alpha}}\mu^{I}
+[\mathcal{M}'{}^{\alpha I}, a'_{\alpha\dot{\alpha}}]\Bigr)
\notag\\
&\qquad\qquad\ 
{}-i\vec{D}\cdot\vec{\tau}^{\dot{\alpha}}{}_{\dot{\beta}} 
\Bigl(\bar{w}^{\dot{\beta}}w_{\dot{\alpha}}
+\bar{a}' {}^{\dot{\beta} \alpha}a'_{\alpha\dot{\alpha}}
\Bigr)
\biggr].
\end{align}
We note that this effective action \eqref{insteff2} is different from 
the fractional D3/D$(-1)$ action in the R-R 3-form background 
at $\mathcal{O}(C\bar{C})$ 
which is obtained as \cite{BiFrFuLe} 
\begin{align}
S^{(0)}_\mathrm{str}&= \frac{2\pi^{2}}{\kappa}\mathrm{tr}_{k}\biggl[
-2\Bigl([\bar{\chi},a'_{\mu}] - \bar{C}_{\mu\nu}a^{\prime \nu}\Bigr)
\Bigl([\chi,a^{\prime \mu}] - C^{\mu\rho}a'_{\rho}\Bigr)\notag\\
&\quad{}+(\bar{\chi}\bar{w}^{\dot{\alpha}}-\bar{w}^{\dot{\alpha}}\bar{\phi})
(w_{\dot{\alpha}}\chi-\phi w_{\dot{\alpha}})
+(\chi\bar{w}^{\dot{\alpha}}-\bar{w}^{\dot{\alpha}}\phi)
(w_{\dot{\alpha}}\bar{\chi}-\bar{\phi}w_{\dot{\alpha}})\notag\\
&\quad{}-i\frac{\sqrt{2}}{2}\bar{\mu}^{I}\epsilon_{IJ}(\mu^{J}\bar{\chi}
-\bar{\phi}\mu^{J})
-i\frac{\sqrt{2}}{4}\mathcal{M}^{\prime\alpha I}\epsilon_{IJ}
\Bigl([\bar{\chi},\mathcal{M}_{\alpha}^{\prime J}]
-\bar{C}_{(\alpha\beta)}\mathcal{M}^{\prime\beta J}\Bigr)
\biggr]
\notag\\
&\quad{}
+S_{\mathrm{ADHM}}.
\label{insteffstring}
\end{align}
The difference between \eqref{insteffstring} and \eqref{insteff2} is
\begin{align}
S^{(0)}_{\mathrm{str}}-S^{(0)}_{\mathrm{eff}}&=
-\frac{\pi^{2}}{\kappa}\mathrm{tr}_{k}\Bigl[
C^{\mu\nu}\bar{C}_{\mu\rho}(\delta_{\nu}{}^{\rho}
\bar{w}^{\dot{\alpha}}w_{\dot{\alpha}}
+4a'_{\nu}a^{\prime \rho})
\Bigr]
\notag\\
&=
-\frac{\pi^{2}}{\kappa}\mathrm{tr}_{k}\Bigl[
C^{\mu\nu}\bar{C}_{\mu\nu}(
\bar{w}^{\dot{\alpha}}w_{\dot{\alpha}}
+a'_{\rho}a^{\prime \rho})
\Bigr]. \label{diff}
\end{align}
Here we have used the relation from the self-duality of 
$C^{\mu\nu}$ and $\bar{C}^{\mu\nu}$ 
\begin{eqnarray}
C_{\mu \rho} \bar{C}_{\nu} {}^{\rho} + C_{\nu \rho} \bar{C}_{\mu} 
{}^{\rho} = \frac{1}{2} C_{\rho \sigma} \bar{C}^{\rho \sigma} 
\delta_{\mu \nu}.
\label{self-dual_formula}
\end{eqnarray}
In order to recover the 
effective action of the D3/D$(-1)$-branes \eqref{insteffstring} 
from the (S,A)-deformed 
super Yang-Mills at $\mathcal{O} (C \bar{C})$, we find that 
the term
\begin{eqnarray}
\delta \mathcal{L} = 
- \frac{g^2}{16 \kappa} 
C^{\rho \sigma} \bar{C}_{\rho 
\sigma} |x|^2 \mathrm{Tr} \left[ F^{\mu \nu} F_{\mu \nu} \right] 
\label{discrepancy}
\end{eqnarray}
needs to be added to the space-time Lagrangian (\ref{N2Lagrangian}).
The contribution $\delta S^{(0)}_{\mathrm{eff}}$ 
to the instanton effective action 
coming from (\ref{discrepancy}) is evaluated 
and coincides with \eqref{diff}
(see appendix \ref{calc} for detail). 
Then 
$S^{(0)}_{\mathrm{eff}} + \delta S^{(0)}_{\mathrm{eff}}$
completely agrees with $S^{(0)}_{\mathrm{str}}$. 
We note that the term (\ref{discrepancy}) which contains space-time 
coordinates explicitly 
cannot be calculated in our previous paper 
\cite{ItNaSa} in which we have treated the constant R-R background only. 
We also note that the term \eqref{discrepancy} does not change 
the self-dual equation at the leading order 
\eqref{N2_leading_self-dual}--\eqref{leading4}.
Hence when we start from the improved space-time Lagrangian 
$\mathcal{L}+\delta \mathcal{L}$, 
we find the same self-dual solution 
\eqref{sol_gauge}--\eqref{sol_adjoint2} and obtain \eqref{insteffstring} 
as the instanton effective action of the improved theory. 

The evaluation of the instanton effective action 
in string theory 
is based on the D$(-1)$-brane effective action. In the presence of the R-R 
background, the modulus $a'_{\mu}$ is stabilized at the origin due to the 
$\mathcal{O} (C \bar{C})$ contribution in (\ref{insteffstring})
which is regarded as the mass term of $a'_{\mu}$.
This moduli stabilization breaks translational invariance in the D3-brane 
world-volume. 
However, from the viewpoint of the D3-brane effective 
action, namely (S,A)-deformed super Yang-Mills theory, 
the background does not induce any terms violating translational symmetry .
This is the reason why there is no $a'_{\mu}$ mass term $\bar{C}^{\mu \nu} 
C_{\nu \rho} a'_{\mu} a^{\prime \rho}$ in the field theory calculation 
in (\ref{insteffstring}).

The action \eqref{insteffstring} 
is invariant under the following deformed supersymmetry transformation 
\begin{align}
\delta a'_{\alpha \dot{\alpha}} &= i \bar{\xi}_{\dot{\alpha} I} 
\mathcal{M}'_{\alpha} {}^{I}, 
&
\delta \mathcal{M}'_{\alpha} {}^I &= - 2 \sqrt{2} \epsilon^{IJ} 
\bar{\xi}^{\dot{\alpha}} {}_J [a'_{\alpha \dot{\alpha}}, \chi] 
+ 2 \sqrt{2}  \bar{\xi}^{\dot{\alpha} I} (\sigma^{\mu})_{\alpha 
\dot{\alpha}} C^{\mu \nu} a'_{\nu}, 
\notag\\
\delta w_{\dot{\alpha}} &= i \bar{\xi}_{\dot{\alpha} I} \mu^I, 
&
\delta \mu^I &= - 2 \sqrt{2} \epsilon^{IJ} \bar{\xi}^{\dot{\alpha}} 
{}_J ( w_{\dot{\alpha}} \chi - \phi w_{\dot{\alpha}}), 
\notag\\
\delta \chi &= 0, 
&
\delta \bar{\chi} &= - \sqrt{2} i \epsilon^{IJ} \bar{\xi}_{\dot{\alpha} I} 
\bar{\psi}^{\dot{\alpha}} {}_J, 
\notag\\
\delta \vec{D} &= - \sqrt{2} \vec{\tau}^{\dot{\alpha}} {}_{\dot{\beta}} 
\bar{\xi}_{\dot{\alpha}}{}^{I} [\bar{\psi}^{\dot{\beta}} {}_I, 
\chi],
&
\delta \bar{\psi}^{\dot{\alpha}} {}_I &= 2 [\chi, \bar{\chi}] 
\bar{\xi}^{\dot{\alpha}} {}_I - i 
\vec{D} \cdot \vec{\tau}^{\dot{\alpha}} {}_{\dot{\beta}} 
\bar{\xi}^{\dot{\beta}} {}_I,
\label{susyeff}
\end{align}
when $C^{\mu \rho} \bar{C}_{\rho\nu} = \bar{C}^{\mu \rho} C_{\rho \nu}$. 
As we will see in next section, this condition is equivalent to the flatness of the $\Omega$-background.
After the topological twist, the above symmetry becomes the 
BRST symmetry 
which is important to apply the localization technique 
\cite{Ne, LoMaNe, NeOk, Shadchin, NaYo} for the calculation of 
the prepotential. 
The instanton effective action is BRST-exact as shown in \cite{BiFrFuLe}.

\section{Relation to the $\Omega$-background deformation}
In the previous section, we showed that the instanton effective 
action in the (S,A)-deformed $\mathcal{N} = 2$ super Yang-Mills theory 
coincides with the D3/D$(-1)$-brane effective action with (S,A)-type background 
if we introduce the additional term (\ref{discrepancy}).
In the following, we discuss 
the relation between 
the $\Omega$-background deformation 
and the (S,A)-deformation of the $\mathcal{N} = 2$ super Yang-Mills theory.

The four-dimensional $\Omega$-deformed $\mathcal{N} = 2$ super 
Yang-Mills Lagrangian $\mathcal{L} (\Omega, \bar{\Omega})$ is 
obtained by 
the dimensional reduction of six-dimensional $\mathcal{N} = 1$ super 
Yang-Mills theory in 
the $\Omega$-background metric \cite{Ne2}
\begin{eqnarray}
ds_6^2 = 2 d z d \bar{z} + (d x^{\mu} + \bar{\Omega}^{\mu} dz + 
\Omega^{\mu} d \bar{z})^2,
\end{eqnarray}
where $z = \frac{1}{\sqrt{2}} (x^5 - i x^6),$ $\bar{z} = \frac{1}{\sqrt{2}} 
(x^5 + i x^6)$.
$\Omega^{\mu}$ and $\bar{\Omega}^{\mu}$ are defined by 
$\Omega^{\mu} \equiv \Omega^{\mu \nu} x_{\nu} 
$, $\bar{\Omega}^{\mu} \equiv \bar{\Omega}^{\mu \nu} x_{\nu} $ 
with constant anti-symmetric matrices 
$\Omega^{\mu \nu} = - \Omega^{\nu \mu}$ and $\bar{\Omega}^{\mu \nu} = - 
\bar{\Omega}^{\nu \mu}$. 
In this background, all nonzero components in the Riemann tensor 
are proportional to $\Omega_{\mu \nu} \bar{\Omega}^{\nu} {}_{\rho} - 
\bar{\Omega}_{\mu \nu} \Omega^{\nu} {}_{\rho}$. 
Then $\Omega$ and $\bar{\Omega}$ are taken to be commutative matrices 
so that space-time is flat.
As we will see, under the identification (\ref{omega_sa}), 
this flatness condition becomes the supersymmetry invariance of the 
instanton effective action.

The six dimensional $\mathcal{N} = 1$ super Yang-Mills action is 
\begin{eqnarray}
S = \int \! d^6 x \ \sqrt{- g} \mathrm{Tr}
\left[
- \frac{1}{4} g^{MP} g^{NQ} F_{MN} F_{PQ} - \frac{i}{2} \bar{\Psi} 
e^{M} {}_m \Gamma^m \mathcal{D}_{M} \Psi
\right],
\end{eqnarray}
where $M, N = 0, \cdots 5$ stands for curved indices in six dimensional 
space-time and $m$ is a local Lorentz index. $e^M {}_m$ is a vielbein 
and $\Gamma^m$ is a six dimensional gamma matrix.
The covariant derivative is defined by $\mathcal{D}_M = D_M - \frac{1}{2} \omega_{M, mn} \Gamma^{mn}$ 
where $D_M$ is an ordinary gauge covariant derivative and 
$\omega_{M,mn}$ is a spin connection. The field strength is defined by 
$F_{MN} = \partial_M A_N - \partial_N A_M + ig [A_M, A_N]$ and $\Psi$ is 
a six dimensional Dirac spinor.
After the dimensional reduction and the Wick rotation, 
we obtain the four-dimensional Lagrangian
\begin{eqnarray}
\mathcal{L} (\Omega, \bar{\Omega}) = \mathcal{L}_0 + \delta \mathcal{L} 
(\Omega, \bar{\Omega}),
\label{OmegaLag}
\end{eqnarray}
where $\mathcal{L}_0$ is the  $\mathcal{N} = 2$ super Yang-Mills Lagrangian 
(\ref{N2SYM}) and $ \delta \mathcal{L} (\Omega, \bar{\Omega})$ is 
\begin{eqnarray}
\delta \mathcal{L} (\Omega, \bar{\Omega}) 
&=&
\frac{1}{k} \mathrm{Tr} \left[ \frac{}{} g F_{\mu \nu} D^{\mu} \bar{\varphi} 
\Omega^{\nu} +  g F_{\mu \nu} D^{\mu} \varphi 
\overline{\Omega}^{\nu} \right. \nonumber \\
& & \qquad + i  g^2 D_{\mu} \bar{\varphi} [\varphi, \bar{\varphi}]
\Omega^{\mu} + i  g^2 D_{\mu} \varphi [\varphi, \bar{\varphi}] 
\overline{\Omega}^{\mu}- g^2 F_{\mu \rho} F_{\nu} {}^{\rho} \Omega^{\mu} \overline{\Omega}^{\nu} 
+ \frac{g^2}{2} D_{\mu} \bar{\varphi} D_{\nu} \bar{\varphi} 
\Omega^{\mu} \Omega^{\nu} 
\nonumber \\
& & \qquad  - g^2  D_{\mu} \bar{\varphi} 
D_{\nu} \varphi \Omega^{\mu} \overline{\Omega}^{\nu} + 
 \frac{g^2}{2} D_{\mu} \varphi D_{\nu} \varphi \overline{\Omega}^{\mu} 
\overline{\Omega}^{\nu} + i  g^3 [\varphi, \bar{\varphi}] F_{\mu \nu} \overline{\Omega}^{\mu} \Omega^{\nu}
 \nonumber \\
& & \qquad - \frac{g}{\sqrt{2}} \Lambda^{\alpha I} D_{\mu} \Lambda_{\alpha I} \overline{\Omega}^{\mu}
- \frac{g}{\sqrt{2}} \overline{\Lambda}_{\dot{\alpha} I} D_{\mu} 
\overline{\Lambda}^{\dot{\alpha} I} \Omega^{\mu} \nonumber \\
& & \left. \qquad - \frac{g}{\sqrt{2}} \Lambda_{\alpha} {}^I 
 \Lambda_{\beta I} \overline{\Omega}^{(\alpha \beta)} - \frac{g}{\sqrt{2}} 
\overline{\Lambda}_{\dot{\alpha} I} \overline{\Lambda}_{\dot{\beta}} 
{}^{I} \Omega^{(\dot{\alpha} \dot{\beta})} \right] 
+ \mathcal{O} (\Omega^3, \overline{\Omega}^3). \label{omega_N2_SYM}
\end{eqnarray}
Here $\bar{\Omega}^{(\alpha \beta)} = \frac{1}{2} \varepsilon^{\alpha 
\gamma} (\sigma_{\mu \nu})_{\gamma}{}^{\beta}\Omega^{\mu\nu},\ 
\bar{\Omega}^{(\dot{\alpha} 
\dot{\beta})} = \frac{1}{2} \varepsilon^{\dot{\alpha} \dot{\gamma}} 
(\bar{\sigma}_{\mu 
\nu})^{\dot{\beta}}{}_{\dot{\gamma}}\bar{\Omega}^{\mu\nu}$, 
$\bar{\sigma}_{\mu \nu} = \frac{1}{4} (\bar{\sigma}^{\mu} \sigma^{\nu} - 
\bar{\sigma}^{\nu} \sigma^{\mu})$.
This part $\delta \mathcal{L} (\Omega, \bar{\Omega})$ can be interpreted as 
a shift of the scalar fields $(\varphi, \bar{\varphi}) \to (\varphi - i 
\Omega^{\mu} D_{\mu}, \bar{\varphi} + i \bar{\Omega}^{\mu} D_{\mu})$ and the modification of 
the complex coupling constant $\tau \to \tau - \frac{4\sqrt{2} i}{g^2} 
\theta_{\alpha} \bar{\theta}_{\beta} \bar{\Omega}^{(\alpha \beta)}$ 
in the $\mathcal{N} = 2$ superfield formalism of super Yang-Mills action \cite{Shadchin}.

Notice that 
once we assume the self-duality condition of $\bar{\Omega}^{\mu\nu}$, 
the term which 
contains $\bar{\Omega}^{(\dot{\alpha} \dot{\beta})}$ vanish 
due to the anti-self-duality of $\bar{\sigma}_{\mu \nu}$.
In the following, we assume self-duality of $\Omega^{\mu \nu}$ and $\bar{\Omega}^{\mu \nu}$.

The Lagrangian \eqref{OmegaLag} contains many 
$x^{\mu}$-dependent interactions and 
looks like quite different from the (S,A)-deformed theory. 
However, the leading order equations of motion for the self-dual 
instanton (\ref{N2SD1})-(\ref{N2SD5}) 
turn out to be not so different from the (S,A)-deformed theory. 
It is given by
\begin{eqnarray}
& & \nabla^2 \varphi^{(0)} - (\nabla^{\mu} F^{(0)}_{\mu \nu})
\Omega^{\nu} + F^{(0)}_{\mu \nu} \Omega^{\mu \nu} 
+ \sqrt{2} i \Lambda^{(0)\alpha I} \Lambda^{(0)}_{\alpha I} = 0, 
\label{omega_eom1} \\
& & \nabla^2 \bar{\varphi}^{(0)} - (\nabla^{\mu} F^{(0)}_{\mu \nu})
\overline{\Omega}^{\nu} + F^{(0)}_{\mu \nu} 
\overline{\Omega}^{\mu \nu} = 0, \label{omega_eom2} \\
& & \nabla^{\mu} (F^{(0)}_{\mu \nu} + \tilde{F}^{(0)}_{\mu \nu}) = 0, 
\label{omega_eom3} \\
& & F^{(0)}_{\mu \nu} = \tilde{F}^{(0)}_{\mu \nu}, \label{omega_eom4} \\
& & i (\sigma^{\mu})_{\alpha \dot{\beta}} \nabla_{\mu} 
\overline{\Lambda}^{(0)\dot{\beta}} {}_I + \sqrt{2} i [\bar{\varphi}^{(0)}, 
\Lambda^{(0)}_{\alpha I}] - \sqrt{2} \Lambda^{(0)}_{\beta I} 
\overline{\Omega}_{\alpha} {}^{\beta} + \frac{1}{\sqrt{2}} \nabla_{\mu} 
\Lambda^{(0)}_{\alpha I} \overline{\Omega}^{\mu} = 0, \label{omega_eom5} 
\\
& & i (\bar{\sigma}^{\mu})^{\dot{\alpha} \alpha} \nabla_{\mu} \Lambda^{(0)}_{\alpha 
} {}^I = 0. \label{omega_eom6}
\end{eqnarray}
The Bianchi identity and the self-dual condition 
$F^{(0)}_{\mu \nu} = \tilde{F}^{(0)}_{\mu \nu}$ can be used to remove the 
second term in (\ref{omega_eom1}) and (\ref{omega_eom2}). 
After the identification
\begin{eqnarray}
\Omega_{\mu \nu} = i C_{\mu \nu}, \ \bar{\Omega}_{\mu \nu} = i \bar{C}_{\mu \nu},
\label{omega_sa}
\end{eqnarray}
the leading order equations of motion 
(\ref{omega_eom1})--(\ref{omega_eom6}) 
agree with the equations 
(\ref{N2_leading_self-dual})-(\ref{leading_eq}) for (S,A)-deformed super 
Yang-Mills theory except for the equation of motion for $\bar{\Lambda}$.
However the contribution 
of $\bar{\Lambda}^{(0)}_{I}$ is just the subleading order 
in $g$, hence it does not contribute to the instanton effective action. 
The $\mathcal{O} (g^0)$ terms in the 
instanton effective action (\ref{omega_N2_SYM}) is given by
\begin{eqnarray}
S^{(0)}_{\mathrm{eff}} (\Omega, \overline{\Omega}) &=& 
\frac{1}{\kappa} \int \! d^4 x \ \mathrm{Tr} \left[ 
 \nabla_{\mu} \varphi^{(0)} \nabla^{\mu} \bar{\varphi}^{(0)} - \frac{i}{\sqrt{2}}
\Lambda^{(0)I} [\bar{\varphi}, \Lambda^{(0)}_{I}] -
\bar{\varphi}^{(0)} F^{(0)}_{\mu \nu} \Omega^{\mu \nu} 
 \right.
\nonumber \\
& & \qquad \left. \qquad - \varphi^{(0)} F^{(0)}_{\mu \nu} \overline{\Omega}^{\mu \nu}
+ \frac{1}{\sqrt{2}} \Lambda^{(0)}_{\alpha} {}^{I} 
\Lambda^{(0)}_{\beta I} \overline{\Omega}^{(\alpha \beta)}  \right] 
\nonumber \\
& & + \frac{1}{\kappa} \int \! d^4 x \ \mathrm{Tr} \left[ \frac{}{}
 F^{(0)}_{\mu \rho} F_{\nu}^{(0)} {}^{\rho} \Omega^{\mu} \overline{\Omega}^{\nu}
+ \frac{1}{\sqrt{2}} \Lambda^{(0) \alpha I} \nabla_{\mu} 
\Lambda^{(0)}_{\alpha I} \bar{\Omega}^{\mu}  \right]. \label{omega_inst}
\end{eqnarray}
The last term in (\ref{omega_inst}) can be rewritten as
\begin{eqnarray}
 \frac{1}{\sqrt{2}} \mathrm{Tr}\Bigl[
\overline{\Omega}^{\mu} \Lambda^{(0)\alpha I} 
\nabla_{\mu} \Lambda^{(0)}_{\alpha I}\Bigr] 
=
\frac{1}{\sqrt{2}}\bar{\Omega}^{\mu\nu}\mathrm{Tr}\Bigl[
\Lambda^{(0)I}\sigma_{\mu\nu}\Lambda^{(0)}_{I}-\partial^{\rho}\bigl(x_{\nu}
\Lambda^{(0)I}\sigma_{\mu\rho}\Lambda^{(0)}_{I}\bigr)\Bigr].
\end{eqnarray}
Here the second term in the right hand side 
is the total derivative and does not contribute to the effective 
action in the instanton background 
since $x_{\nu}\Lambda^{(0)I}\sigma_{\mu\rho}\Lambda^{(0)}_{I}$
behaves as $|x|^{-5}$ for large $|x|$. 
Therefore we have
\begin{eqnarray}
S^{(0)}_{\mathrm{eff}} (\Omega, \overline{\Omega}) &=& 
\frac{1}{\kappa} \int \! d^4 x \ \mathrm{Tr} \left[ 
\nabla_{\mu} \varphi^{(0)} \nabla^{\mu} \bar{\varphi}^{(0)} -\frac{i}{\sqrt{2}}
\Lambda^{(0)I} [\bar{\varphi}, \Lambda^{(0)}_{I}] -
\bar{\varphi}^{(0)} F^{(0)}_{\mu \nu} \Omega^{\mu \nu} 
 \right.
\nonumber \\
& & \qquad \left. \qquad - \varphi^{(0)} F^{(0)}_{\mu \nu} \overline{\Omega}^{\mu \nu}
- \frac{1}{\sqrt{2}} \Lambda^{(0)}_{\alpha} {}^I 
\Lambda^{(0)}_{\beta I} \overline{\Omega}^{(\alpha \beta)}  \right] \nonumber\\
& & + \frac{1}{\kappa} \int \! d^4 x \ \mathrm{Tr} \left[ \frac{}{}
F^{(0)}_{\mu \rho} F_{\nu}^{(0)} {}^{\rho} \Omega^{\mu} \overline{\Omega}^{\nu}
\right].
\label{omegainst}
\end{eqnarray}
This result coincides with the one obtained from the improved action 
discussed in section 2.
The last term in (\ref{omegainst}) agrees with (\ref{discrepancy}) 
by using the relation (\ref{self-dual_formula})
and self-duality of $F^{(0)}_{\mu\nu}$.

We note that we compared the space-time action deformed in the R-R 3-form background
with the action in the $\Omega$-background without the R-symmetry gauge field Wilson line.
If one includes the R-symmetry gauge field Wilson line, one gets the topological field 
theory in the $\Omega$-background \cite{Ne, LoMaNe} which differs from 
the $\mathcal{N} = 2$ action in the same background by topological terms \cite{Shadchin}. 
Therefore the instanton effective action remains the same by the twisting.
The deformed action is BRST-exact and the instanton effective action is 
also written in the BRST-exact form.
Although it is not clear at this moment how to introduce the R-symmetry gauge Wilson line
 in the fractional
D3-branes, the BRST transformations would correspond to the deformed supersymmetry 
transformation in the R-R 3-form background.

\section{Conclusions and discussions}
In this paper, we investigate (anti-)self-dual solutions in the deformed 
$\mathcal{N} = 2$ super Yang-Mills theory. 
The theory is realized on the (fractional) D3-branes at the 
fixed point of the orbifold $\mathbf{C} \times \mathbf{C}^2/\mathbf{Z}_2$ 
in the presence of the R-R 3-form field strength background.
The R-R 3-form background $\mathcal{F}^{(\alpha \beta ) [AB]}$ is scaled as $(2 \pi \alpha')^{\frac{1}{2}}
\mathcal{F}^{(\alpha \beta) [AB]} = \textrm{fixed}$ in order to give the deformation 
parameters $C, \bar{C}$ the mass dimension one. 
In the $\mathcal{N} = 2$ supergravity context, these are interpreted as the graviphoton and 
the vector backgrounds, respectively \cite{BiFrFuLe}. 

The instanton solution is expressed 
in terms of the the ADHM moduli and 
the deformation parameters. With this solution, we explicitly evaluate 
the instanton effective action for the self-dual solution 
using the field theoretical method.
The result agrees with the one previously obtained in the string theory 
calculation \cite{BiFrFuLe} up to the first order in the deformation 
parameters but differs from that at $\mathcal{O} (C \bar{C})$.
However, once we add the translational symmetry breaking term 
to the (S,A)-deformed action and consider the improved action, 
we obtain the string theory result.

The deformed $\mathcal{N} = 2$ instanton effective 
action derived from the improved action is the same with the the action in $\Omega$-background 
\cite{Ne} despite the fact that the space-time action has a 
different form. 
The instanton effective action is invariant under deformed supersymmetry 
if $C$ and $\bar{C}$ commute with each other, which corresponds to the flatness condition of the $\Omega$-background.

It is interesting to consider the deformation in the (A,S)-type background.
In \cite{ItNaSa}, we have shown that the (A,S)-type R-R 3-form 
background $\mathcal{F}^{[\alpha \beta] (AB)}$ 
induces mass terms for the chiral fermion $\Lambda$ and other adjoint 
scalar field interactions\footnote{
In \cite{ItNaSa}, the mass terms for the anti-chiral fermion 
$\bar{\Lambda}$ was considered. Here we consider different chirality of the (A,S)-type 
background considered in \cite{ItNaSa} to generate the mass term for $\Lambda$.
}.
In self-dual case, we can show that the 
bosonic interactions induced by the (A,S)-background are 
sub-leading order and do not contribute to the instanton effective action.
The only relevant part is the mass term for $\Lambda$ which 
contributes to the equation of motion of $\bar{\Lambda}$.
However, as we have seen section 2, the 
solution of $\bar{\Lambda}$ does not contribute to the instanton effective 
action because it enters in the space-time action as the sub-leading 
part. Therefore the only modification in the instanton effective action by the (A,S)-background is 
just the mass term of the $\Lambda$ which can be easily evaluated by 
Corrigan's inner product formula (\ref{cor}). 
Similar to the (S,A)-type deformation, 
there are no (A,S)-background corrections 
to the instanton effective action 
for the anti-self-dual case because corrections are sub-leading order.

It is possible to generalize the results in this paper to 
$\mathcal{N} = 4$ and $\mathcal{N} = 2^{*}$ super Yang-Mills theories.
These generalizations will appear in a forthcoming paper \cite{ItNaSaSa}.

\subsection*{Acknowledgments}
S.~S. would like to thank Claus Montonen for reading the manuscript.
The work of K.~I. is  supported in part by the Grant-in-Aid for Scientific 
Research from Ministry of Education, Science, 
Culture and Sports of Japan. 
The work of H.~N. is the result of research activities (Astrophysical 
Research Center for the Structure and Evolution of the Cosmos 
(ARCSEC)) and grant No.\ R01-2006-000-10965-0 from the Basic 
Research Program supported by KOSEF. 
The work of S.~S. is supported by bilateral exchange program between 
Japan Society for the Promotion of Science (JSPS) and the Academy of 
Finland (AF). 

\begin{appendix}
\section{The ADHM construction in (deformed) $\mathcal{N} = 2$ supersymmetric Yang-Mills Theory}\label{revADHM}
Here we briefly summarize the ADHM construction \cite{AtHiDrMa, DoHoKhMa}. 
As we have seen in the equations of motion 
\eqref{N2_leading_self-dual}, \eqref{leading4} 
for the (S,A)-deformed action, 
the self-dual equations for the gauge field and the spinor field do not 
change in the deformed theory.
Therefore one can solve them by the ADHM construction 
based on the undeformed theory.
We introduce the $(N+2k)\times 2k$ matrix $\Delta_{\lambda j\dot{\alpha}}$ 
which is given by 
\begin{equation}
\Delta_{\lambda j\dot{\alpha}}
=a_{\lambda j\dot{\alpha}}+b_{\lambda j}{}^{\beta}
\sigma_{\mu\beta\dot{\alpha}}x^{\mu},
\end{equation}
where $\lambda=1,2,\ldots,N+2k$ and $i,j=1,2,\ldots,k$. 
$k$ is the instanton number. 
$a_{\lambda j\dot{\alpha}}$ and $b_{\lambda j}{}^{\beta}$ are 
the constant matrices. They are decomposed as
\begin{equation}
a_{\lambda j\dot{\alpha}}=\binom{w_{uj\dot{\alpha}}}
{(a'_{ij})_{\alpha\dot{\alpha}}}, \quad 
b_{\lambda j}{}^{\beta}=\binom{0}{\delta_{ij}\delta_{\alpha}{}^{\beta}}, \quad
\lambda=u+i\alpha,\quad u=1,2,\ldots,N.
\end{equation}
The matrix $\Delta$ should satisfy the following ADHM constraints,
\begin{equation}
\bar{\Delta}^{\dot{\alpha}\lambda}_{i}\Delta_{\lambda j\dot{\beta}}
=(f^{-1})_{ij}\delta^{\dot{\alpha}}{}_{\dot{\beta}},\quad
f=\biggl[\frac{1}{2}\bar{w}^{\dot{\alpha}}w_{\dot{\alpha}}
+(x_{\mu}+a'_{\mu})^{2}\biggr]^{-1}, \quad
a'_{\mu}=\frac{1}{2}\bar{\sigma}_{\mu}^{\dot{\alpha}\alpha}
a'_{\alpha\dot{\alpha}}, 
\label{ADHMapp}
\end{equation}
where $a'_{\mu}$, $w_{\dot{\alpha}}$, and $\bar{w}^{\dot{\alpha}}$ 
are called ADHM moduli. 
The ADHM constraints in terms of $a'_{\alpha \dot{\alpha}}, 
\bar{w}_{\dot{\alpha}}$ are 
\begin{equation}
(\vec{\tau})^{\dot{\alpha}}_{~\dot{\beta}}
(\bar{w}^{\dot{\beta}}w_{\dot{\alpha}}
+\bar{a}^{\prime\dot{\beta}\alpha}a'_{\alpha\dot{\alpha}})=0,\quad 
a'_{\mu}=\bar{a}'_{\mu}. \label{ADHM2}
\end{equation}
We also introduce $(N+2k)\times N$ matrix $U$ which satisfies 
\begin{equation}
\bar{\Delta}U=0,\quad \bar{U}U=\boldsymbol{1}_{N},\quad 
U\bar{U}+\Delta_{\dot{\alpha}} f\bar{\Delta}^{\dot{\alpha}}=
\boldsymbol{1}_{N+2k},
\end{equation}
where $\boldsymbol{1}_{n}$ is the $n \times n$ identity matrix. 
The self-dual gauge field is constructed from $U$ as
\begin{equation}
A_{\mu}^{(0)}=-i\bar{U}\partial_{\mu}U.
\end{equation}
The corresponding field strength $F_{\mu\nu}^{(0)}$ is 
\begin{equation}
F_{\mu\nu}^{(0)}=-4i\bar{U}b^{\alpha}(\sigma_{\mu\nu})_{\alpha}{}^{\beta}f
\bar{b}_{\beta}U.
\label{SDA}
\end{equation}
The self-duality of $F_{\mu\nu}^{(0)}$ immediately follows from that of 
$\sigma_{\mu\nu}$.

Nextly we consider the fermionic moduli which appear as the fermionic zero 
modes on the instanton background. 
We solve the Dirac equation on the self-dual background 
$\bar{\sigma}^{\mu}\nabla_{\mu}\Lambda^{(0)I}=0$. 
The ansatz of the solution is 
\begin{equation}
\Lambda^{(0)I}_{\alpha}=\Lambda_{\alpha} (\mathcal{M}^I)=\bar{U}
(\mathcal{M}^{I}f\bar{b}_{\alpha}-b_{\alpha}f\bar{\mathcal{M}}^{I})U,
\label{af}
\end{equation}
where $\mathcal{M}^{I}$ is the $(N+2k)\times k$ constant matrix. 
Plugging \eqref{af} to the Dirac equation, we obtain
\begin{equation}
(\bar{\sigma}^{\mu})^{\dot{\alpha}\alpha}\nabla_{\mu}\Lambda^{(0)I}_{\alpha}=
2\bar{U}b^{\alpha}f(\bar{\mathcal{M}}^{I}\Delta^{\dot{\alpha}}
+\bar{\Delta}^{\dot{\alpha}}\mathcal{M}^{I})f\bar{b}_{\alpha}U.
\end{equation}
Then we have the fermionic ADHM constraint 
\begin{equation}
\bar{\mathcal{M}}^{I}\Delta^{\dot{\alpha}}
+\bar{\Delta}^{\dot{\alpha}}\mathcal{M}^{I}=0,
\label{fADHMapp}
\end{equation}
or equivalently
\begin{equation}
\bar{\mu}^{I}w_{\dot{\alpha}}+\bar{w}_{\dot{\alpha}}\mu^{I}
+[\mathcal{M}^{\prime\alpha I},a'_{\alpha\dot{\alpha}}]=0,
\quad \mathcal{M}^{\prime I}_{\alpha}=\bar{\mathcal{M}}^{\prime I}_{\alpha},
\end{equation}
where we have decomposed $\mathcal{M}^{I}$ as
\begin{equation}
\mathcal{M}^{I}_{\lambda j}
=\binom{\mu^{I}_{uj}}{(\mathcal{M}^{\prime I}_{\alpha})_{ij}}.
\end{equation}
$\mathcal{M}^{\prime I}_{\alpha}$, $\mu^{I}$, $\bar{\mu}^{I}$ are 
called fermionic ADHM moduli.

Now we solve the equation of motion of the scalar field $\varphi^{(0)}$ 
\begin{equation}
\nabla^{2}\varphi^{(0)}+i\sqrt{2}\Lambda^{(0)I}\Lambda_{I}^{(0)}
+iC^{\mu\nu}F_{\mu\nu}^{(0)}=0.
\label{scalareq}
\end{equation}
First we consider the case of $C^{\mu\nu}=0$.
The ansatz of the solution \cite{DoHoKhMa} is
\begin{equation}
\varphi^{(0)}=-i\frac{\sqrt{2}}{4}\epsilon_{IJ}\bar{U}\mathcal{M}^{I}f
\bar{\mathcal{M}}^{J}U
+\bar{U}\begin{pmatrix} \phi & 0 \\
 0 & \chi\boldsymbol{1}_{2}
\end{pmatrix}U.
\label{scalaransatz0}
\end{equation}
The asymptotic behavior of \eqref{scalaransatz0} is given by 
$\lim_{|x|\to\infty}\varphi^{(0)}=\phi$. Computing $\nabla^{2}\varphi^{(0)}$, 
one can show that
\begin{align}
\nabla^{2}\varphi^{(0)}
&=
-i\sqrt{2}\epsilon_{IJ}\Lambda^{(0)I}\Lambda^{(0)J}
+4\bar{U}bf\Biggl[i\frac{\sqrt{2}}{4}\epsilon_{IJ}
\bar{\mathcal{M}}^{I}\mathcal{M}^{J}
-\{f^{-1},\chi\}+\bar{\Delta}^{\dot{\alpha}}
\begin{pmatrix} \phi & 0 \\
 0 & \chi\boldsymbol{1}_{2}
\end{pmatrix}
\Delta_{\dot{\alpha}}\Biggr]f\bar{b}U
\notag\\
&=
-i\sqrt{2}\epsilon_{IJ}\Lambda^{(0)I}\Lambda^{(0)J}
+4\bar{U}bf\biggl(i\frac{\sqrt{2}}{4}\epsilon_{IJ}
\bar{\mathcal{M}}^{I}\mathcal{M}^{J}-\boldsymbol{L}\chi
+\bar{w}^{\dot{\alpha}}\phi w_{\dot{\alpha}}\biggr)f\bar{b}U,
\end{align}
where $\boldsymbol{L}\chi$ is defined by
\begin{equation}
\boldsymbol{L}\chi=
\frac{1}{2}\bigl\{\bar{w}^{\dot{\alpha}}w_{\dot{\alpha}},\chi\bigr\}
+\Bigl[a'_{\mu},[a^{\prime \mu},\chi]\Bigr].
\end{equation}
Then $\varphi^{(0)}$ satisfies the equation of motion if $\chi$ satisfies
\begin{equation}
\boldsymbol{L}\chi=i\frac{\sqrt{2}}{4}\epsilon_{IJ}
\bar{\mathcal{M}}^{I}\mathcal{M}^{J}
+\bar{w}^{\dot{\alpha}}\phi w_{\dot{\alpha}}.
\end{equation}
In the case of $C^{\mu\nu}\neq 0$, the ansatz of the solution is changed as \cite{BiFrFuLe}
\begin{equation}
\varphi^{(0)}=-i\frac{\sqrt{2}}{4}\epsilon_{IJ}\bar{U}\mathcal{M}^{I}f
\bar{\mathcal{M}}^{J}U
+\bar{U}\begin{pmatrix} \phi & 0 \\
 0 & \chi\boldsymbol{1}_{2}+\boldsymbol{1}_{k}C 
\end{pmatrix}U,
\end{equation}
where $C$ is the $2\times 2$ matrix of which component is 
$C_{\alpha}{}^{\beta}=
\frac{1}{2}C^{\mu\nu}(\sigma_{\mu\nu})_{\alpha}{}^{\beta}$. 
Now one can show that
\begin{equation}
\begin{split}\label{eom1}
\nabla^{2}\varphi^{(0)}
&=
-i\sqrt{2}\epsilon_{IJ}\Lambda^{(0)I}\Lambda^{(0)J}
+4\bar{U}bf\Biggl[i\frac{\sqrt{2}}{4}\epsilon_{IJ}
\bar{\mathcal{M}}^{I}\mathcal{M}^{J}\\
&\qquad
{}-2f^{-1}C-\{f^{-1},\chi\}+\bar{\Delta}^{\dot{\alpha}}
\begin{pmatrix} \phi & 0 \\
 0 & \chi\boldsymbol{1}_{2}+\boldsymbol{1}_{k}C 
\end{pmatrix}
\Delta_{\dot{\alpha}}\Biggr]f\bar{b}U.
\end{split}
\end{equation}
The third term in 
the right hand side of \eqref{eom1} becomes the deformation term in 
the equation of motion \eqref{scalareq} due to \eqref{SDA}.
The $C$-dependent part in the last term is
\begin{equation}
\bar{\Delta}^{\dot{\alpha}}
\begin{pmatrix} 0 & 0 \\
 0 & \boldsymbol{1}_{k}C 
\end{pmatrix}
\Delta_{\dot{\alpha}}
=(\bar{a}'+\bar{x})^{\dot{\alpha}\alpha}C_{\alpha}{}^{\beta}
(a'+x)_{\beta\dot{\alpha}}
=C^{\mu\nu}[a'_{\mu},a'_{\nu}].
\end{equation}
Then we obtain
\begin{equation}
\begin{split}
\nabla^{2}\varphi^{(0)}
&=
-i\sqrt{2}\epsilon_{IJ}\Lambda^{(0)I}\Lambda^{(0)J}-iC^{\mu\nu}F^{(0)}_{\mu\nu}
\\
&\qquad
+4\bar{U}bf\biggl(i\frac{\sqrt{2}}{4}\epsilon_{IJ}^{}
\bar{\mathcal{M}}^{I}\mathcal{M}^{J}
-\boldsymbol{L}\chi
+\bar{w}^{\dot{\alpha}}\phi w_{\dot{\alpha}}
+C^{\mu\nu}[a'_{\mu},a'_{\nu}]
\biggr)f\bar{b}U.
\end{split}
\end{equation}
Hence $\varphi^{(0)}$ is the solution of the deformed equation of motion
if $\chi$ satisfies
\begin{equation}
\boldsymbol{L}\chi=i\frac{\sqrt{2}}{4}\epsilon_{IJ}
\bar{\mathcal{M}}^{I}\mathcal{M}^{J}
+\bar{w}^{\dot{\alpha}}\phi w_{\dot{\alpha}}
+C^{\mu\nu}[a'_{\mu},a'_{\nu}].
\end{equation}
We can also solve the instanton equation of $\bar{\varphi}^{(0)}$
in a similar way.

\section{Detailed calculations of the instanton effective action\label{calc}}
\subsection{Calculation of \eqref{secondthird} and \eqref{fourth}}
Here we give the detail for the calculation of 
\eqref{secondthird} and \eqref{fourth}.
In order to calculate \eqref{secondthird}, 
we use the formula
\begin{equation}
\bigl[\bar{\varphi}^{(0)},\Lambda_{\alpha I}^{(0)}\bigr]
=(\sigma^{\mu})_{\alpha\dot{\alpha}}\nabla_{\mu}\bar{\psi}^{\dot{\alpha}}_{I}
+\Lambda_{\alpha}(\mathcal{N}_{I})
+\epsilon_{IJ} \bar{C}_{\alpha}{}^{\beta} \Lambda_{\beta}(\mathcal{M}^J),
\label{n2id}
\end{equation}
where $\bar{\psi}_{I}$ and $\mathcal{N}_{I}$ are given by
\begin{align}
\bar{\psi}_{I}^{\dot{\alpha}}
&=
\bar{\psi}_{I}^{(1)\dot{\alpha}}+\bar{\psi}_{I}^{(2)\dot{\alpha}},
\\[2mm]
\bar{\psi}_{I}^{(1)\dot{\alpha}}
&=
\frac{1}{2}\bar{U}\left[ - \mathcal{M}_{I}f
\bar{\Delta}^{\dot{\alpha}}
\begin{pmatrix} \bar{\phi} & 0 \\[2mm]
0 & \bar{\chi}\boldsymbol{1}_{2}+\boldsymbol{1}_{k}\bar{C} \end{pmatrix}
+
\begin{pmatrix} \bar{\phi} & 0 \\[2mm]
0 & \bar{\chi}\boldsymbol{1}_{2}+\boldsymbol{1}_{k}\bar{C} \end{pmatrix}
\Delta^{\dot{\alpha}}f\bar{\mathcal{M}}_{I}\right]U,
\\[2mm]
\bar{\psi}_{I}^{(2)\dot{\alpha}}
&=
\bar{U}\begin{pmatrix} 0 & 0 \\[2mm]
0 & \mathcal{G}_{I}^{\dot{\alpha}}\boldsymbol{1}_{2}
\end{pmatrix}U,\quad \partial_{\mu}\mathcal{G}_{I}^{\dot{\alpha}}=0,
\\[2mm]
\mathcal{N}_{I}
&=
\begin{pmatrix} \bar{\phi} & 0 \\[2mm]
0 & \bar{\chi}\boldsymbol{1}_{2}+\boldsymbol{1}_{k}\bar{C} \end{pmatrix}
\mathcal{M}_{I}-\mathcal{M}_{I}\bar{\chi}
+2
\begin{pmatrix} 0 & 0 \\[2mm]
0 & \mathcal{G}_{I}^{\dot{\alpha}} \end{pmatrix}
a_{\dot{\alpha}}-2a_{\dot{\alpha}}\mathcal{G}_{I}^{\dot{\alpha}}.
\end{align}
Here 
the $k\times k$ matrix $\mathcal{G}_{I}^{\dot{\alpha}}$ 
is chosen such that 
$\mathcal{N}_{I}$ 
satisfies the fermionic ADHM condition 
$\bar{\mathcal{N}}_{I}\Delta^{\dot{\alpha}}
+\bar{\Delta}^{\dot{\alpha}}\mathcal{N}_{I}=0$.
From the formula \eqref{n2id}, \eqref{secondthird} becomes 
\begin{align}
&\int\!d^{4}x\,\frac{1}{\kappa}\mathrm{Tr}\biggl[
-\frac{i}{\sqrt{2}}
\Lambda^{(0)\alpha I}\bigl[\bar{\varphi}^{(0)},\Lambda_{\alpha I}^{(0)}\bigr]
-\frac{i}{\sqrt{2}}\bar{C}^{(\alpha\beta)}\Lambda_{\alpha}^{(0)I}
\Lambda_{\beta I}^{(0)}\biggr]\notag\\
&\qquad\qquad=
\int\!d^{4}x\,\frac{1}{\kappa}\mathrm{Tr}\biggl[-\frac{i}{\sqrt{2}}
\partial_{\mu}\bigl(\Lambda(\mathcal{M}^{I})\sigma^{\mu}\bar{\psi}_{I}\bigr)
-\frac{i}{\sqrt{2}}
\Lambda^{\alpha}(\mathcal{M}^{I})\Lambda_{\alpha}(\mathcal{N}_{I})
\biggr].\label{n2st}
\end{align}
The first term in the right hand side 
of \eqref{n2st} vanishes since 
$\Lambda(\mathcal{M}^{I})\sigma^{\mu}\bar{\psi}_{I}$ behaves as 
$|x|^{-5}$ for large $|x|$. 
The second term can be evaluated using Corrigan's inner-product formula 
\cite{DoKhMa, DoHoKhMaVa}
\begin{align}
\int\!d^{4}x\,\frac{1}{\kappa}\mathrm{Tr}\Bigl[
\Lambda^{\alpha}(\mathcal{M}^{I})\Lambda_{\alpha}(\mathcal{N}_{I})\Bigr]
&=
- \frac{\pi^{2}}{2 \kappa} \mathrm{tr}_{k}\Bigl[\bar{\mathcal{M}}^{I}
(\mathcal{P}_{\infty}+1)\mathcal{N}_{I}
+\bar{\mathcal{N}}_{I}(\mathcal{P}_{\infty}+1)\mathcal{M}^{I}\Bigr]
\notag\\
&= -
\frac{2\pi^{2}}{\kappa} \epsilon_{IJ}\mathrm{tr}_{k}\biggl[
\bar{\mu}^{I}\bar{\phi}\mu^{J}
-\bar{\mathcal{M}}^{I}\mathcal{M}^{J}\bar{\chi}
+\frac{1}{2}\bar{C}^{(\alpha\beta)}\mathcal{M}^{\prime I}_{\alpha}
\mathcal{M}^{\prime J}_{\beta}\biggr],
\label{cor}
\end{align}
where $\mathcal{P}_{\infty}=\lim_{|x|\to\infty}U\bar{U}$. 
Since the part proportional to $\mathcal{G}_{I}^{\dot{\alpha}}$ vanishes in 
\eqref{cor} by fermionic ADHM condition \eqref{fADHM}, we do not need to solve 
$\mathcal{G}_{I}^{\dot{\alpha}}$ explicitly. 
Then we obtain \eqref{secondthird} 
\begin{align}
&\int\!d^{4}x\,\frac{1}{\kappa}\mathrm{Tr}\biggl[
-\frac{i}{\sqrt{2}}
\Lambda^{(0)\alpha I}\bigl[\bar{\varphi}^{(0)},\Lambda_{\alpha I}^{(0)}\bigr]
- \frac{i}{\sqrt{2}}\bar{C}^{(\alpha\beta)}\Lambda_{\alpha}^{(0)I}
\Lambda_{\beta I}^{(0)}\biggr]\notag\\
&\qquad\qquad=
\frac{1}{\kappa} \sqrt{2}\pi^{2}i \epsilon_{IJ}
\mathrm{tr}_{k}\biggl[\bar{\mu}^{I}\bar{\phi}\mu^{J}
-\bar{\mathcal{M}}^{I}\mathcal{M}^{J}\bar{\chi}
+\frac{1}{2}\bar{C}^{(\alpha\beta)}\mathcal{M}^{\prime I}_{\alpha}
\mathcal{M}^{\prime J}_{\beta}\biggr].
\label{second}
\end{align}

Nextly we prove \eqref{fourth}.
The left hand side in \eqref{fourth} can be rewritten as
\begin{align}
&\int\!d^{4}x\,\frac{1}{\kappa}\mathrm{Tr}\Bigl[
- i \bar{\varphi}^{(0)} F^{(0)}_{\mu \nu} {C}^{\mu \nu}
\Bigr]
\notag\\
&\qquad\qquad=
-8C_{\alpha}{}^{\beta}
\!\int\! d^{4}x\,\frac{1}{\kappa}\mathrm{Tr}\left[\bar{U}
\begin{pmatrix}
 \bar{\phi} & 0 \\
 0 & \bar{\chi}\boldsymbol{1}_{2}+\boldsymbol{1}_{k}\bar{C}
 \end{pmatrix}
\mathcal{P}b^{\alpha}f\bar{b}_{\beta}U
\right]
\notag\\
&\qquad\qquad=
-2C_{\alpha}{}^{\beta}\!\int\! d^{4}x\,\frac{1}{\kappa}\mathrm{Tr}
\Bigl[
(\sigma^{\mu\nu})_{\beta}{}^{\alpha}\nabla_{\mu}R_{\nu}
+2\bar{U}b^{\alpha}\{f,\bar{\chi}\}\bar{b}_{\beta}U
\Bigr].
\label{fourth_a}
\end{align}
Here $R_{\mu}$ is defined by
\begin{equation}
R_{\mu}=\frac{1}{2}\bar{U}
\left[b^{\alpha}(\sigma_{\mu})_{\alpha\dot{\alpha}}f\bar{\Delta}^{\dot{\alpha}}
\begin{pmatrix}
 \bar{\phi} & 0 \\
 0 & \bar{\chi}\boldsymbol{1}_{2}+\boldsymbol{1}_{k}\bar{C}
 \end{pmatrix}
-
\begin{pmatrix}
 \bar{\phi} & 0 \\
 0 & \bar{\chi}\boldsymbol{1}_{2}+\boldsymbol{1}_{k}\bar{C}
 \end{pmatrix}
\Delta_{\dot{\alpha}}(\bar{\sigma}_{\mu})^{\dot{\alpha}\alpha}f\bar{b}_{\alpha}
\right]U.
\end{equation}
Then the first term of \eqref{fourth_a} is evaluated as
\begin{align}
-2C_{\alpha}{}^{\beta}\!\int\! d^{4}x\,\frac{1}{\kappa}\mathrm{Tr}
\Bigl[
(\sigma^{\mu\nu})_{\beta}{}^{\alpha}\nabla_{\mu}R_{\nu}
\Bigr]
&=
2C^{\mu\nu}\!\int\! d^{4}x\,\frac{1}{\kappa}\partial_{\mu}\mathrm{Tr}R_{\nu}
\notag\\
&= \frac{1}{\kappa}
\pi^{2}C^{\mu\nu}\bar{C}_{\mu\nu}
\mathrm{tr}_{k}\bigl[\bar{w}^{\dot{\alpha}}w_{\dot{\alpha}}\bigr].
\label{fourth1}
\end{align}
The second term of \eqref{fourth_a} becomes a total derivative 
and is calculated as
\begin{align}
- 4C_{\alpha}{}^{\beta}\!\int\! d^{4}x\,\frac{1}{\kappa}\mathrm{Tr}\Bigl[
\bar{U}b^{\alpha}\{f,\bar{\chi}\}\bar{b}_{\beta}U
\Bigr]
&=\frac{4}{\kappa}C^{\mu\nu}\!\int\! d^{4}x\,\partial_{\mu}\mathrm{tr}_{k}
\Bigl[\bigl[f,a'_{\nu}\bigr]\bar{\chi}\Bigr]
\notag\\
&=- \frac{4\pi^{2}}{\kappa}\mathrm{tr}_{k}\Bigl[
C^{\mu\nu}[a'_{\mu},a'_{\nu}]\bar{\chi}\Bigr],
\label{fourth2}
\end{align}
where we have used the asymptotic behavior of $f$ given by 
\begin{equation}
f=\frac{1}{|x|^{2}}\boldsymbol{1}_{k}
-\frac{2x^{\lambda}}{|x|^{4}}a'_{\lambda}+\mathcal{O}(|x|^{-4}).
\end{equation}
Finally, we obtain the result in \eqref{fourth}
\begin{align}
&\int\!d^{4}x\,\frac{1}{\kappa}\mathrm{Tr}\Bigl[
- i \bar{\varphi}^{(0)} F^{(0)}_{\mu \nu} {C}^{\mu \nu}
\Bigr]=
\frac{\pi^{2}}{\kappa}\mathrm{tr}_{k}\Bigl[
-4C^{\mu\nu}[a'_{\mu},a'_{\nu}]\bar{\chi}
+C^{\mu\nu}\bar{C}_{\mu\nu}
\bar{w}^{\dot{\alpha}}w_{\dot{\alpha}}
\Bigr].
\end{align}

\subsection{Verification of the form of the discrepancy term}
The contribution from \eqref{discrepancy} to the instanton effective action , 
$\delta S^{(0)}_{\mathrm{eff}}$ is 
\begin{equation}
\delta S^{(0)}_{\mathrm{eff}}
=
- \frac{1}{16 \kappa} \int \! d^4 x \ 
C^{\rho \sigma} \bar{C}_{\rho \sigma} |x|^2 \mathrm{Tr} 
\left[ F^{(0)\mu \nu} F^{(0)}_{\mu \nu} \right].
\label{add2}
\end{equation}
{}From Osborn's formula \cite{Os} $\mathrm{Tr} 
\left[ F^{(0)\mu \nu} F^{(0)}_{\mu \nu} \right]=-\Box^2 \mathrm{tr}_k \log f$, 
$\delta S^{(0)}_{\mathrm{eff}}$ can be rewritten in a total derivative as 
\begin{eqnarray}
\delta S^{(0)}_{\mathrm{eff}}
&=& \frac{1}{16 \kappa} C^{\rho \sigma} \bar{C}_{\rho \sigma} \int \! d^4 x \ 
|x|^2 \Box^2 \mathrm{tr}_k \log f \nonumber \\
&=& \frac{1}{16 \kappa} C^{\rho \sigma} \bar{C}_{\rho \sigma} \int \! d^4 x \ 
\partial_{\mu} \left[ 
|x|^2 \Box \partial^{\mu} - 2 x^{\mu} \Box + 8 \partial^{\mu} 
\right] \mathrm{tr}_k \log f.
\label{additional}
\end{eqnarray}
Plugging the explicit form of $f$ \eqref{ADHMapp} into \eqref{additional}, 
we obtain \eqref{diff}. 

\end{appendix}

\end{document}